\documentclass[aps,prl,preprint,tightenlines,superscriptaddress,showpacs,byrevtex,subfigure]{revtex4}

\usepackage{graphicx}
\usepackage{color}

\begin{document}
\begin{flushright}
{BELLE-CONF-0652}\\
\end{flushright}

\title{ \quad\\[0.5cm] 
Search for Lepton Flavor
Violating $\tau^-$ 
{Decays}\\
into $\ell^-\eta$,  $\ell^-\eta'$ and  $\ell^-\pi^0$ 
}

\affiliation{Budker Institute of Nuclear Physics, Novosibirsk}
\affiliation{Chiba University, Chiba}
\affiliation{Chonnam National University, Kwangju}
\affiliation{University of Cincinnati, Cincinnati, Ohio 45221}
\affiliation{University of Frankfurt, Frankfurt}
\affiliation{The Graduate University for Advanced Studies, Hayama} 
\affiliation{Gyeongsang National University, Chinju}
\affiliation{University of Hawaii, Honolulu, Hawaii 96822}
\affiliation{High Energy Accelerator Research Organization (KEK), Tsukuba}
\affiliation{Hiroshima Institute of Technology, Hiroshima}
\affiliation{University of Illinois at Urbana-Champaign, Urbana, Illinois 61801}
\affiliation{Institute of High Energy Physics, Chinese Academy of Sciences, Beijing}
\affiliation{Institute of High Energy Physics, Vienna}
\affiliation{Institute of High Energy Physics, Protvino}
\affiliation{Institute for Theoretical and Experimental Physics, Moscow}
\affiliation{J. Stefan Institute, Ljubljana}
\affiliation{Kanagawa University, Yokohama}
\affiliation{Korea University, Seoul}
\affiliation{Kyoto University, Kyoto}
\affiliation{Kyungpook National University, Taegu}
\affiliation{Swiss Federal Institute of Technology of Lausanne, EPFL, Lausanne}
\affiliation{University of Ljubljana, Ljubljana}
\affiliation{University of Maribor, Maribor}
\affiliation{University of Melbourne, Victoria}
\affiliation{Nagoya University, Nagoya}
\affiliation{Nara Women's University, Nara}
\affiliation{National Central University, Chung-li}
\affiliation{National United University, Miao Li}
\affiliation{Department of Physics, National Taiwan University, Taipei}
\affiliation{H. Niewodniczanski Institute of Nuclear Physics, Krakow}
\affiliation{Nippon Dental University, Niigata}
\affiliation{Niigata University, Niigata}
\affiliation{University of Nova Gorica, Nova Gorica}
\affiliation{Osaka City University, Osaka}
\affiliation{Osaka University, Osaka}
\affiliation{Panjab University, Chandigarh}
\affiliation{Peking University, Beijing}
\affiliation{University of Pittsburgh, Pittsburgh, Pennsylvania 15260}
\affiliation{Princeton University, Princeton, New Jersey 08544}
\affiliation{RIKEN BNL Research Center, Upton, New York 11973}
\affiliation{Saga University, Saga}
\affiliation{University of Science and Technology of China, Hefei}
\affiliation{Seoul National University, Seoul}
\affiliation{Shinshu University, Nagano}
\affiliation{Sungkyunkwan University, Suwon}
\affiliation{University of Sydney, Sydney NSW}
\affiliation{Tata Institute of Fundamental Research, Bombay}
\affiliation{Toho University, Funabashi}
\affiliation{Tohoku Gakuin University, Tagajo}
\affiliation{Tohoku University, Sendai}
\affiliation{Department of Physics, University of Tokyo, Tokyo}
\affiliation{Tokyo Institute of Technology, Tokyo}
\affiliation{Tokyo Metropolitan University, Tokyo}
\affiliation{Tokyo University of Agriculture and Technology, Tokyo}
\affiliation{Toyama National College of Maritime Technology, Toyama}
\affiliation{University of Tsukuba, Tsukuba}
\affiliation{Virginia Polytechnic Institute and State University, Blacksburg, Virginia 24061}
\affiliation{Yonsei University, Seoul}
  \author{K.~Abe}\affiliation{High Energy Accelerator Research Organization (KEK), Tsukuba} 
  \author{K.~Abe}\affiliation{Tohoku Gakuin University, Tagajo} 
  \author{I.~Adachi}\affiliation{High Energy Accelerator Research Organization (KEK), Tsukuba} 
  \author{H.~Aihara}\affiliation{Department of Physics, University of Tokyo, Tokyo} 
  \author{D.~Anipko}\affiliation{Budker Institute of Nuclear Physics, Novosibirsk} 
  \author{K.~Aoki}\affiliation{Nagoya University, Nagoya} 
  \author{T.~Arakawa}\affiliation{Niigata University, Niigata} 
  \author{K.~Arinstein}\affiliation{Budker Institute of Nuclear Physics, Novosibirsk} 
  \author{Y.~Asano}\affiliation{University of Tsukuba, Tsukuba} 
  \author{T.~Aso}\affiliation{Toyama National College of Maritime Technology, Toyama} 
% ...
\collaboration{The Belle Collaboration}

\begin{abstract}
We have searched for lepton flavor violating $\tau$ 
{decays} 
with a pseudoscalar meson 
{($\eta$, $\eta'$ and $\pi^0$)}
using a data sample of 401 fb$^{-1}$ collected 
{with} 
the Belle detector at the
KEKB asymmetric-energy $e^+e^-$ collider. 
No evidence for these decays is found {and} 
we set the following upper limits 
{on} the branching fractions:  
${\cal{B}}(\tau^-\rightarrow e^-\eta) < 9.2\times 10^{-8}$, 
${\cal{B}}(\tau^-\rightarrow \mu^-\eta) < 6.5\times 10^{-8}$,
${\cal{B}}(\tau^-\rightarrow e^-\eta') < 1.6\times 10^{-7}$, 
${\cal{B}}(\tau^-\rightarrow \mu^-\eta') < 1.3\times 10^{-7}$  
${\cal{B}}(\tau^-\rightarrow e^-\pi^0) < 8.0\times 10^{-8}$
and 
${\cal{B}}(\tau^-\rightarrow \mu^-\pi^0) < 1.2\times 10^{-7}$    
at the 90\% confidence level, respectively. 
These results improve the previously published limits 
by factors from 2.3 to 6.4.
\end{abstract}
\pacs{11.30.Fs; 13.35.Dx; 14.60.Fg}
\maketitle
 \section{Introduction}

{Lepton flavor violation (LFV)
{is allowed} in many extensions of the Standard Model (SM),
{such as} Supersymmetry (SUSY) and leptoquark models.}
{In particular,} lepton flavor violating decays
with a pseudoscalar meson ($M^0 = \eta, \eta'$ and $\pi^0$)
are discussed
{in  models with} 
Higgs-mediated LFV processes~\cite{cite:higgs},
heavy singlet Dirac neutrinos~\cite{cite:amon},
$R-$parity violation in SUSY~\cite{cite:rpv, cite:rpv2},
dimension-six effective fermionic operators that induce $\tau-\mu$
mixing~\cite{cite:six_fremionic}
{and others~\cite{Li:2005rr}.
The best upper limits  
{for these} 
modes are {in the range}
{(1.5$-$10)~$\times~10^{-7}$}
at the 90\% confidence level.
{These were} obtained
{by}  the Belle experiment}
using 154 fb${}^{-1}$ of data~\cite{cite:leta}.

In this paper,
we {report}  a {search} for
{lepton flavor violating  decays}
{with a}  pseudoscalar meson  
$\tau^-\rightarrow \ell^-M^0$
($\ell = e \mbox{ or } \mu$
and $M^0 = \eta, \eta'\mbox{ or } \pi^0$ )\footnotemark[2]
{using  401 fb$^{-1}$ of data
collected at the $\Upsilon(4S)$ resonance
and 60 MeV below it}
with the Belle detector at the KEKB  $e^+e^-$
asymmetric-energy collider~\cite{kekb}.
\footnotetext[2]{Unless otherwise stated, charge
conjugate decays are
{included}
throughout
this paper.}

The Belle detector is a large-solid-angle magnetic spectrometer that
consists of a silicon vertex detector (SVD),
a 50-layer central drift chamber (CDC),
an array of aerogel threshold \v{C}erenkov counters (ACC), 
a barrel-like arrangement of
time-of-flight scintillation counters (TOF), 
and an electromagnetic calorimeter
{comprised of}
CsI(Tl) {crystals (ECL), all located} inside
a superconducting solenoid coil
that provides a 1.5~T magnetic field.
An iron flux-return located outside of the coil is instrumented to 
detect $K_{\rm{L}}^0$ mesons
and to identify muons (KLM).
The detector is described in detail elsewhere~\cite{Belle}.

{Particle identification
is very important in this measurement.
{We use particle identification
likelihood variables based on}}
the ratio of the energy
deposited in the ECL to the momentum measured in the SVD and CDC,
the shower shape in the ECL,
the particle range in the KLM,
the hit information from the ACC,
{the measured $dE/dX$} in the CDC
and {the {particle} time-of-flight} from the TOF.
{{For lepton identification,
we {form} likelihood ratios ${\cal P}(e)$~\cite{EID} 
and ${\cal P}({\mu})$~\cite{MUID}
based on the}
electron and  muon probabilities, respectively,
{which are}
determined by
the responses of the appropriate subdetectors.}

In order to determine the event selection 
{requirements},
we use the Monte Carlo (MC) samples.
{The following MC programs have been} 
used to
generate background events:
KORALB/TAUOLA~\cite{cite:koralb_tauola} for $\tau^+\tau^-$,
QQ~\cite{cite:qq} for $B\bar{B}$ and continuum,
BHLUMI~\cite{BHLUMI} for {Bhabha events,}
KKMC~\cite{KKMC} for $e^+e^-\rightarrow\mu^+\mu^-$ and
AAFH~\cite{AAFH} for two-photon processes.
Signal MC is generated by KORALB/TAUOLA.
{Signal $\tau$ decays are  two-body
and assumed}
to  have a uniform angular distribution
{in the $\tau$} lepton's rest frame.
{All kinematic variables are
calculated in the laboratory frame
{unless otherwise specified.}
In particular,
variables
calculated in the $e^+e^-$ center-of-mass (CM) frame
are indicated by the superscript ``CM''.}
\section{Event Selection}

{We search for $\tau^+\tau^-$ events
in which one $\tau$ 
decays
{into a lepton and a pseudoscalar meson} 
{on} the signal side,
while 
the other $\tau$ 
decays 
into  one charged track 
with a {sign} opposite to that of {the}
signal-side lepton
and any number of additional photons and neutrinos 
on the tag side.}
Thus, the experimental signature is:
\begin{center}
$\left\{
\tau^- \rightarrow \ell^-(=e^-\mbox{ or }\mu^-) + M^0 (=\eta,
\eta'\mbox{ or }\pi^0 )
\right\} 
~+
~ \left\{ \tau^+ \rightarrow ({\rm a~track})^+ + (n^{\rm TAG}_{\gamma} \ge 0)
 + X(\rm{missing}) 
\right\}$.
\end{center}
We reconstruct a
pseudoscalar meson
in the following modes:
$\eta\to\gamma\gamma$ and $\pi^+\pi^-\pi^0(\to\gamma\gamma)$,
$\eta'\to\rho(\to\pi^+\pi^-)\gamma$ and $\eta(\to\gamma\gamma)\pi^+\pi^-$,
$\pi^0\to\gamma\gamma$.
While the $\pi^0\to\gamma\gamma$ and $\eta\to\gamma\gamma$ modes
correspond to the 1-1 prong configuration, the other modes   
{give  3-1} prong configurations.
{All charged tracks} and photons 
are required to be reconstructed 
{within {the} fiducial volume,} 
defined by $-0.866 < \cos\theta < 0.956$,
where $\theta$ is the polar angle with
{respect to the direction opposite to the $e^+$ beam.}
We select charged tracks with
{momenta} 
{transverse to the $e^+$ beam}
$p_t > 0.1$ GeV/$c$
{while 
the photon  energies must satisfy
the requirement 
$E_{\gamma} > 0.1$ GeV} 
($0.05$ GeV) 
{for the 1-1 prong
(the 3-1 prong) configuration.}

%
%

{Candidate $\tau$-pair events are required to have} 
two and four tracks
{with}  a zero net charge,
for {the} 
1-1 and 3-1 prong 
{configurations,} respectively.
{Event particles {are} separated into two 
hemispheres referred to as {the} signal and 
tag {sides}
using the plane perpendicular to the thrust
axis~\cite{thrust}. 
Whereas the tag side
contains a single track,
the signal side contains one or three tracks.}
For {the} 1-1 prong configuration, 
we require that 
{the number of photons {on} the signal side
be two or three.}
The track 
{on} the signal side
is required to satisfy 
the lepton identification selection.
{The electron and muon {identification} criteria are}
${\cal P}(e) > 0.9$ with $p >$ 0.7 GeV/c
and 
${\cal P}(\mu) > 0.9$ with $p >$ 0.7 GeV/c,
respectively.
{The efficiencies 
{for} electron and muon 
{identification after} these requirements
are 92\% and 88\%, repectively.}
{To {reduce} fake 
pseudoscalar meson {candidates,} 
we reject radiative photons 
from {electrons on} the signal side if $\cos \theta_{e\gamma}>$ 0.99.}

To ensure that the missing particles are neutrinos rather
than photons or charged particles
{that fall outside the detector acceptance,}
we impose additional requirements on the missing
momentum vector, $\vec{p}_{\rm miss}$,
calculated by subtracting the
vector sum of the momenta
of
all tracks and photons
from the sum of the $e^+$ and $e^-$ beam momenta.
We require that the magnitude of $\vec{p}_{\rm miss}$
{be} greater than
0.4 GeV/$c$ and that
{its direction point into}
the fiducial volume of the
detector.
{Since neutrinos are emitted only on the tag side,
the direction of
{$\vec{p}_{\rm miss}$
should lie within the tag side of the event.}}
The cosine of the
opening angle between
{$\vec{p}_{\rm miss}$}
and
{the} thrust axis ({on} the signal side)
in the CM system,
{$\cos \theta^{\mbox{\rm \tiny CM}}_{\rm tag-thrust}$,}
{is therefore required to be less than {$-0.55$}}.

\subsection{Event selection for 
{the} $\eta\to\gamma\gamma$ mode}

The $\eta$ meson is
reconstructed by
combining {two photons.}
The mass window 
is chosen to be
0.515 GeV/$c^2$  $< m_{\gamma\gamma} < 0.570$ GeV/$c^2$,
{which corresponds to $-$3.0 and $+$2.5 standard
deviations ($\sigma$)
in {terms} of the mass resolution.}
{To avoid fake $\eta$ candidates, 
we reject
those photons
that  form $\pi^0$ candidates
in association with any other {photon} with $E_{\gamma} > 0.05$ GeV,
within
the $\pi^0$ mass window, 
{0.10 GeV/$c^2$ $< M_{\gamma\gamma} < 0.16$ GeV/$c^2$}.}
To suppress fake $\eta$ events 
{from beam background and initial state radiation (ISR),}  
we require that the higher and lower energy  {photons} 
in an $\eta$ candidate 
($E_{\gamma1}$ and $E_{\gamma2}$)
{satisfy the requirement} 
$E_{\gamma1} > $ 0.6 GeV and $E_{\gamma2} >$ 0.25 GeV, respectively.
{To reduce background} 
from Bhabha and $\mu^+\mu^-$ events,
we require the momentum of a lepton and 
{a tag-side charged particle}
{to be} less than 4.5 {GeV/$c$.}

The total visible energy in the CM frame{,}
{$E^{\mbox{\rm{\tiny{CM}}}}_{\rm{vis}}$,}
is defined as the sum of the energies
{of the $\eta$ candidate,
the lepton,
the tag-side track
(with {a} pion mass hypothesis)
and all photon candidates.
{We require $E^{\mbox{\rm{\tiny{CM}}}}_{\rm{vis}}$
to satisfy}
{the condition:}
5.29 GeV $< E^{\mbox{\rm{\tiny{CM}}}}_{\rm{vis}} < 10.0$ GeV.
To reduce background from $\mu^+\mu^-$, 
{two-photon} and Bhabha events,
{we {further} require $E^{\mbox{\rm{\tiny{CM}}}}_{\rm{vis}}$ 
to satisfy 
{the veto condition:}
$E^{\mbox{\rm{\tiny{CM}}}}_{\rm{vis}} > 8.5$ GeV for
the muon mode (electron mode) 
if the track {on} the tag side is {a} muon (electron).}
The cosine of the
opening angle between 
the lepton
and
{the} $\eta$
in the CM system,
{$\cos \theta^{\mbox{\rm \tiny CM}}_{\rm \ell-\eta}$,}
{is required to 
{lie in the range} 
$0.50 < \cos \theta^{\mbox{\rm \tiny
CM}}_{\rm \ell-\eta} < 0.85$}.
The reconstructed mass {on} 
the tag side using a track 
(with {a} pion mass hypothesis)
and {photons},
$m_{\rm tag}$, 
is 
less than 1.777 GeV/$c^2$.
{In order to suppress background
from $q\bar{q}$ ($q = u, d, s, c$)
continuum events,}
the following requirement
on
the number of photon candidates on the tag side
is imposed:
$n^{\rm{TAG}}\leq 2$.

%
%

The correlation between {the}
momentum of the track on the tag side,
$p^{\rm CM}_{{\rm tag}}$,
and
the cosine of the opening angle
{between the thrust and missing particle,}
$\cos \theta^{\rm CM}_{{\rm thrust-miss}}$
{in the CM system}
{is employed to further suppress
{backgrounds} 
from generic $\tau^+\tau^-$
and $\mu^+\mu^-$ events via the following requirements:}
$p^{\rm CM}_{{\rm tag}} > 1.1\log(\cos \theta^{\rm CM}_{{\rm thrust-miss}}+0.92)+5.5$,
and 
$p^{\rm CM}_{{\rm tag}} < 5\cos \theta^{\rm CM}_{{\rm thrust-miss}}+7.8$
where $p^{\rm CM}_{\rm tag}$ is in GeV/$c$
(see Fig.~\ref{fig:tag-cos}). 
Finally, 
we require
the following relation
between the missing momentum $p_{\rm{miss}}$ and
missing mass squared
$m^2_{\rm{miss}}$ 
{to} 
further suppress background from generic $\tau^+\tau^-$
and
continuum background.
In signal events, 
two neutrinos are included 
if the $\tau$ decay {on} the tag side is leptonic decay,
while 
one neutrino is included if the $\tau$ decay {on} the tag
side is {a} hadronic decay.
{{Therefore,} 
we 
separate events into
{two classes}
according to the track {on} the tag side: leptonic or hadronic.}
{We apply the following requirements 
$p_{\rm miss} > -10m^2_{\rm miss}+4$
and 
$p_{\rm miss} > 1.1m^2_{\rm miss}-0.3$ for a leptonic mode 
on the tag side,
and require
$p_{\rm miss} > -5m^2_{\rm miss}-0.25$
and 
$p_{\rm miss} > 2.1m^2_{\rm miss}-0.3$ for a  hadronic mode 
on the tag side, 
where $p_{{\rm tag}}$ is in GeV/$c$
(see Fig.~\ref{fig:pmiss_vs_mmiss2}). }
{Following all the selection criteria,}
the signal detection efficiencies 
for {the} {$\tau^-\rightarrow e^-\eta(\to\gamma\gamma)$} and
{$\tau^-\rightarrow\mu^-\eta(\to\gamma\gamma)$} modes are
{5.08\% and 7.13\%, respectively.}

\begin{figure}
\begin{center}
\resizebox{0.8\textwidth}{0.8\textwidth}{\includegraphics 
{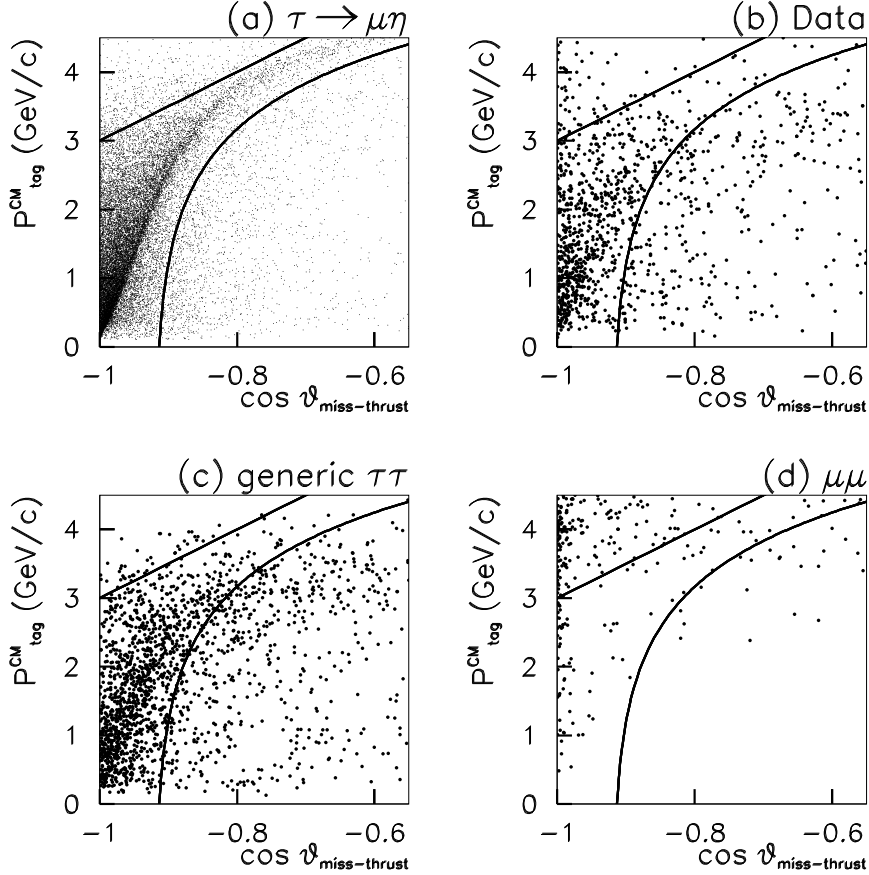}}
 \vspace*{-1cm}
 \caption{
{Scatter-plots 
 of (a) signal MC ($\tau^-\rightarrow\mu^-\eta(\to\gamma\gamma)$),  
(b) data,  
(c) generic $\tau^+\tau^-$ MC events 
 and (d) $\mu^+\mu^-$ MC events on {the}
$p^{\rm CM}_{\rm tag}$ vs $\cos \theta_{\rm miss-thrust}$ plane.
 Selected regions are indicated by {the curves} with arrows.}
 \label{fig:tag-cos}
 }
 \end{center}
\end{figure}

\begin{figure}
\begin{center}
 \resizebox{0.8\textwidth}{0.4\textwidth}{\includegraphics
 {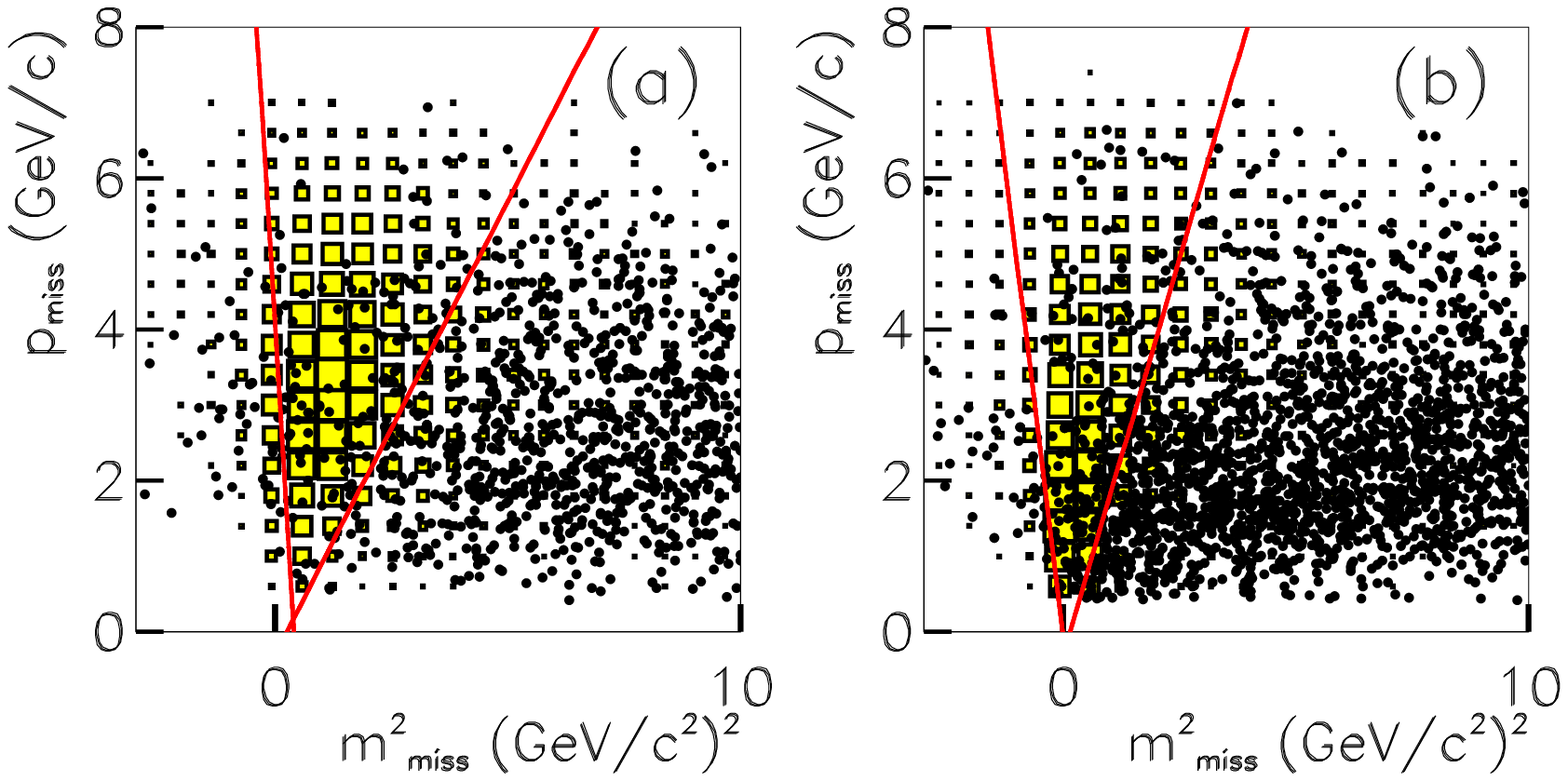}}
 \vspace*{-0.5cm}
 \caption{
{Scatter-plots of missing momentum, 
$p_{\rm miss}$, vs missing mass squared, $m_{\rm miss}^2$,
for (a) leptonic and (b) hadronic tags.}
Selected regions are indicated by lines.
The data are 
the dots.
The filled boxes show the signal MC ($\tau^-\to\mu^-\eta$) distribution
with arbitrary normalization.}
\label{fig:pmiss_vs_mmiss2}
 \end{center}
\end{figure}

\subsection{Event selection for the $\eta\to\pi^+\pi^-\pi^0$ mode}

%
%

The $\eta$ meson is 
reconstructed 
from 
$\pi^+\pi^-\pi^0$.
{The $\pi^0$ candidates are formed from a pair of photons
that satisfy
0.115 GeV/$c^2$ $< M_{\gamma\gamma} < $ 0.152 GeV/$c^2$ {($\pm$2.5$\sigma$).}}
with $p_{\pi^0} >$ 0.1 GeV/c,
where $p_{\pi^0}$ is the $\pi^0$ momentum in the laboratory
system. 
The mass window {for} 
$\eta\to\pi^+\pi^-\pi^0$
is chosen as
0.532 GeV/$c^2$ $< M_{3\pi}< 0.562$ GeV/$c^2$,
{which corresponds to $\pm$3.0$\sigma$.}
Figure~\ref{fig:mueta3pi}
shows the mass distribution {for the}
$\eta\to\pi^+\pi^-\pi^0$ 
{candidates}.
Good agreement between 
data and the MC expectation is observed. 

%
%
\begin{figure}
\begin{center}
 \resizebox{0.4\textwidth}{0.4\textwidth}{\includegraphics
 {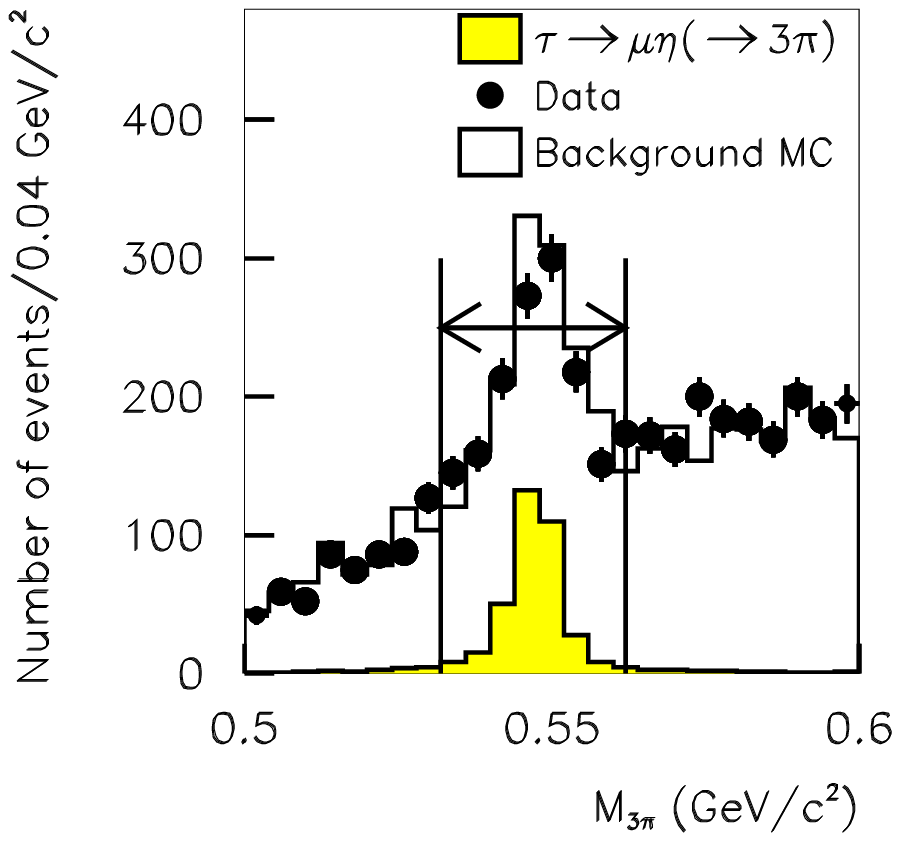}}
\vspace*{-0.5cm}
 \caption{ 
 {Invariant mass distribution of $\eta\to\pi^+\pi^-\pi^0$ candidates.
 While the signal MC ($\tau^-\rightarrow\mu^-\eta(\to\pi^+\pi^-\pi^0)$) 
 distribution is normalized arbitrarily, 
 {the data and background MC} are normalized to the same luminosity.
 {The selected region is indicated  
 by arrows {between} the vertical lines} }
}
\label{fig:mueta3pi}
 \resizebox{0.8\textwidth}{0.4\textwidth}{\includegraphics
 {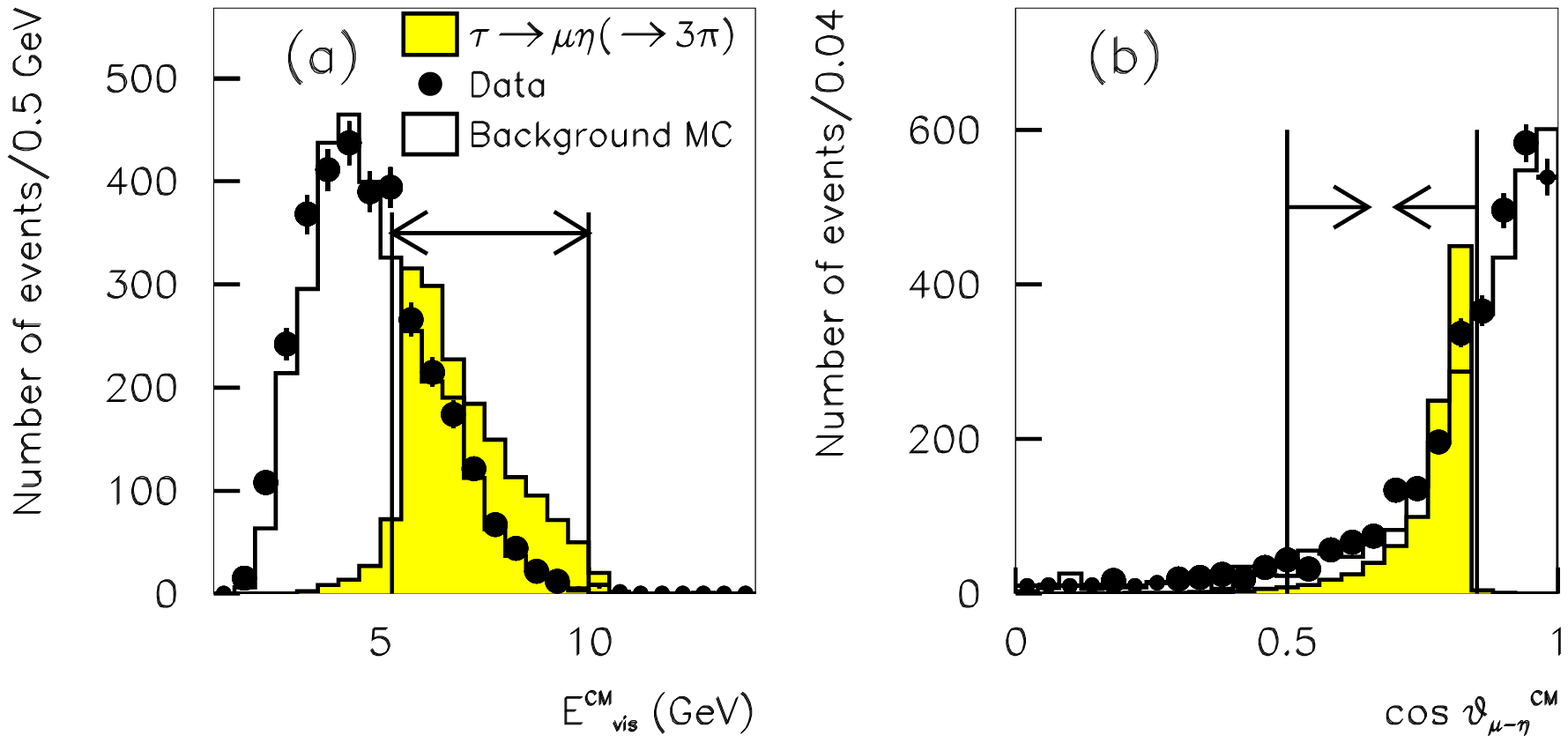}}
\vspace*{-0.5cm}
 \caption{ 
 {Kinematic distributions used in the event selection
 after {the initial} 
 $\eta\to\pi^+\pi^-\pi^0$ selection, muon {identification and
 missing particles requirements:}
 (a) The total visible energy in the CM frame;
 (b) the opening angle between {the} muon and $\eta$ candidate in the CM frame;
 {while} the signal MC ($\tau^-\rightarrow\mu^-\eta(\to\pi^+\pi^-\pi^0)$) 
 distribution is normalized arbitrarily, 
 {the data and background MC} are normalized to the same luminosity.
 Selected regions are indicated  
 {by the arrows between the vertical lines.}}}
\label{fig:cut_3pi}
\end{center}
\end{figure}

%
%

%
%

{Similarly for} the $\eta(\to\gamma\gamma)$ mode,
we require 
the following:
5.29 GeV $< E^{\mbox{\rm{\tiny{CM}}}}_{\rm{vis}} < $10.0 GeV, 
0.50 $< \cos \theta^{\mbox{\rm \tiny CM}}_{\rm \ell-\eta} <$ 0.85
(see Fig.~\ref{fig:cut_3pi} (a) and (b))
and
$m_{\rm {tag}} < $ 1.777 GeV/$c^2$.
The requirement on
the number of the photon candidates for the signal
{is  }
$n^{\rm{SIG}}\leq 1$.
{In addition}
we apply the following {requirements:} 
$p_{\rm miss} > -5m^2_{\rm miss}-0.25$
and 
$p_{\rm miss} > 2.1m^2_{\rm miss}-0.3$ {for hadronic tags,}
and 
$p_{\rm miss} > -10m^2_{\rm miss}-4$
and 
$p_{\rm miss} > 1.1m^2_{\rm miss}-1$ {for leptonic tags,}
respectively. 
{After all the selection criteria,}
the signal detection efficiencies 
for {the} {$\tau^-\rightarrow e^-\eta(\to\pi^+\pi^-\pi^0)$} and
{$\tau^-\rightarrow\mu^-\eta(\to\pi^+\pi^-\pi^0)$} modes are
{5.25\% and 7.60\%, 
respectively.}

\subsection{Event selection for the
$\eta'\to\rho(\to\pi^+\pi^-)\gamma$ mode}

For the  $\rho\to\pi^+\pi^-$ {selection,}
the mass window is chosen to be
0.550 GeV/$c^2$ $< m_{\pi\pi} < 0.900$ GeV/$c^2$.
{We} reconstruct {$\eta'$ candidates} 
using a $\rho$ candidate and {a} photon
on the signal side.
The $\eta'$ mass window is chosen to be
$0.930$ GeV/$c^2$ $< m_{\rho\gamma} < 0.970$ GeV/$c^2$,
{which corresponds to $-3.0$ and $+$2.5$\sigma$.}
Furthermore,
{we veto photons from $\pi^0$ candidates}
in order to avoid fake $\eta'$ candidates
from  $\pi^0\to\gamma\gamma$.
We remove events if a $\pi^0$ with
{invariant mass in the range}
0.10 GeV/$c^2$ $< M_{\gamma\gamma} < 0.16$ GeV/$c^2$
is reconstructed 
by a photon 
from the $\eta'$ candidate 
and another photon  with $E_{\gamma} > 0.05$ GeV.
Figure~\ref{fig:muetap_rhogam}
shows the $\rho\to\pi^+\pi^-$  and
$\eta'\to\rho\gamma$ mass distributions.
{Dominant backgrounds}
for this mode 
come from $\tau^-\to h^-\rho^0\nu_{\tau}(+\pi^0)$ with 
{a photon} from
{$\pi^0$} decay, beam background and {ISR.}
{{As shown in Fig.~\ref{fig:muetap_rhogam},
we see no $\eta'$ peak 
either {in}  data or in  MC}
since 
{decay modes with {an} 
$\eta'$ are very rare}
and {are} not included 
{in the} 
generic $\tau$ decay {model}~\cite{PDG}.}

\begin{figure}
 \resizebox{0.8\textwidth}{0.4\textwidth}{\includegraphics
 {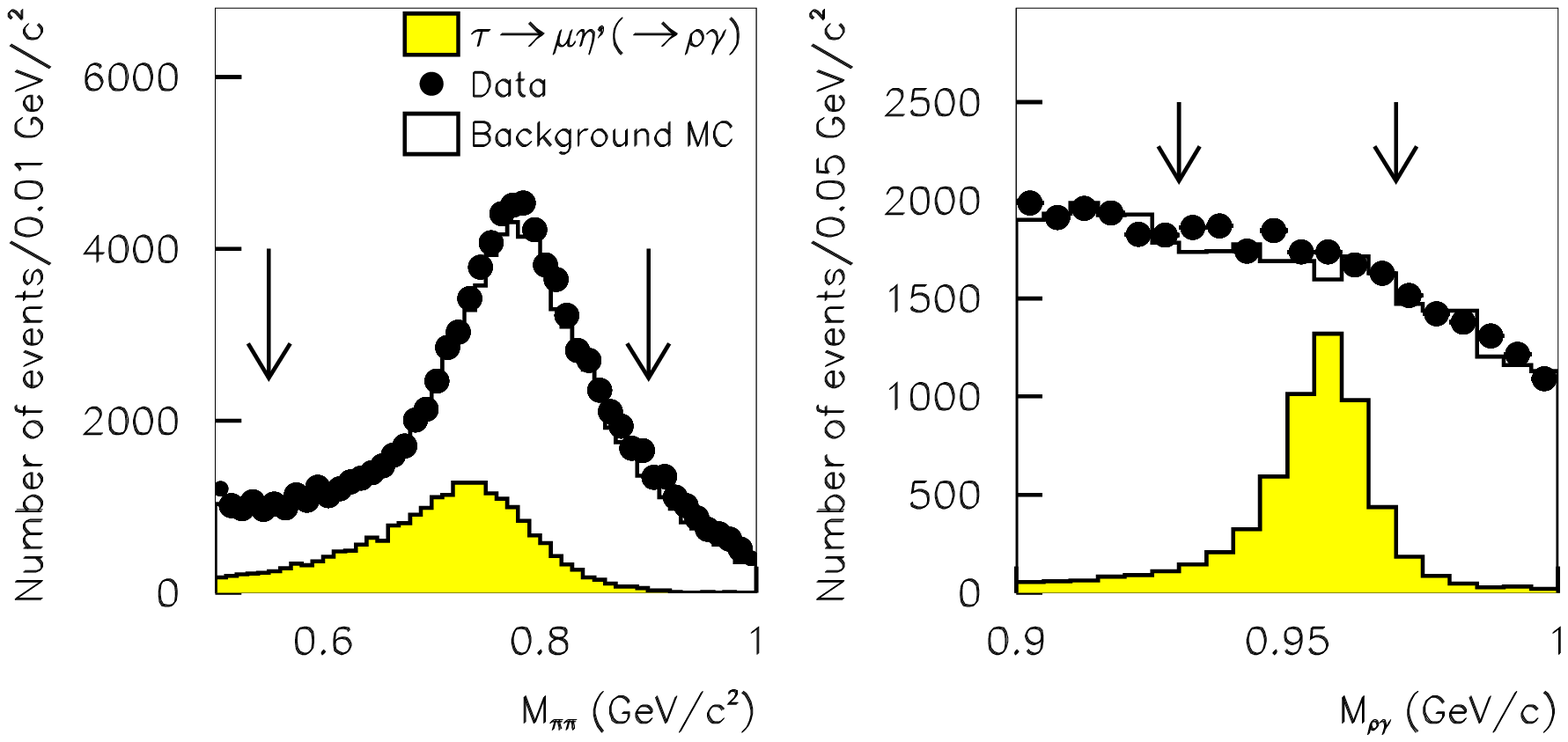}}
 \vspace*{-0.5cm}
\caption{{The $\rho\to\pi^+\pi^-$ (left) and
$\eta'\to\rho\gamma$ (right) mass 
 {distributions.}
 While the signal MC ($\tau^-\rightarrow\mu^-\eta'$) 
 distribution is normalized arbitrarily, 
 {the data and background MC} are normalized to the same luminosity.
 {Selected regions are indicated  
 by {the} arrows.}}}
\label{fig:muetap_rhogam}
\end{figure}

\begin{figure}
 \resizebox{0.8\textwidth}{0.8\textwidth}{\includegraphics
   {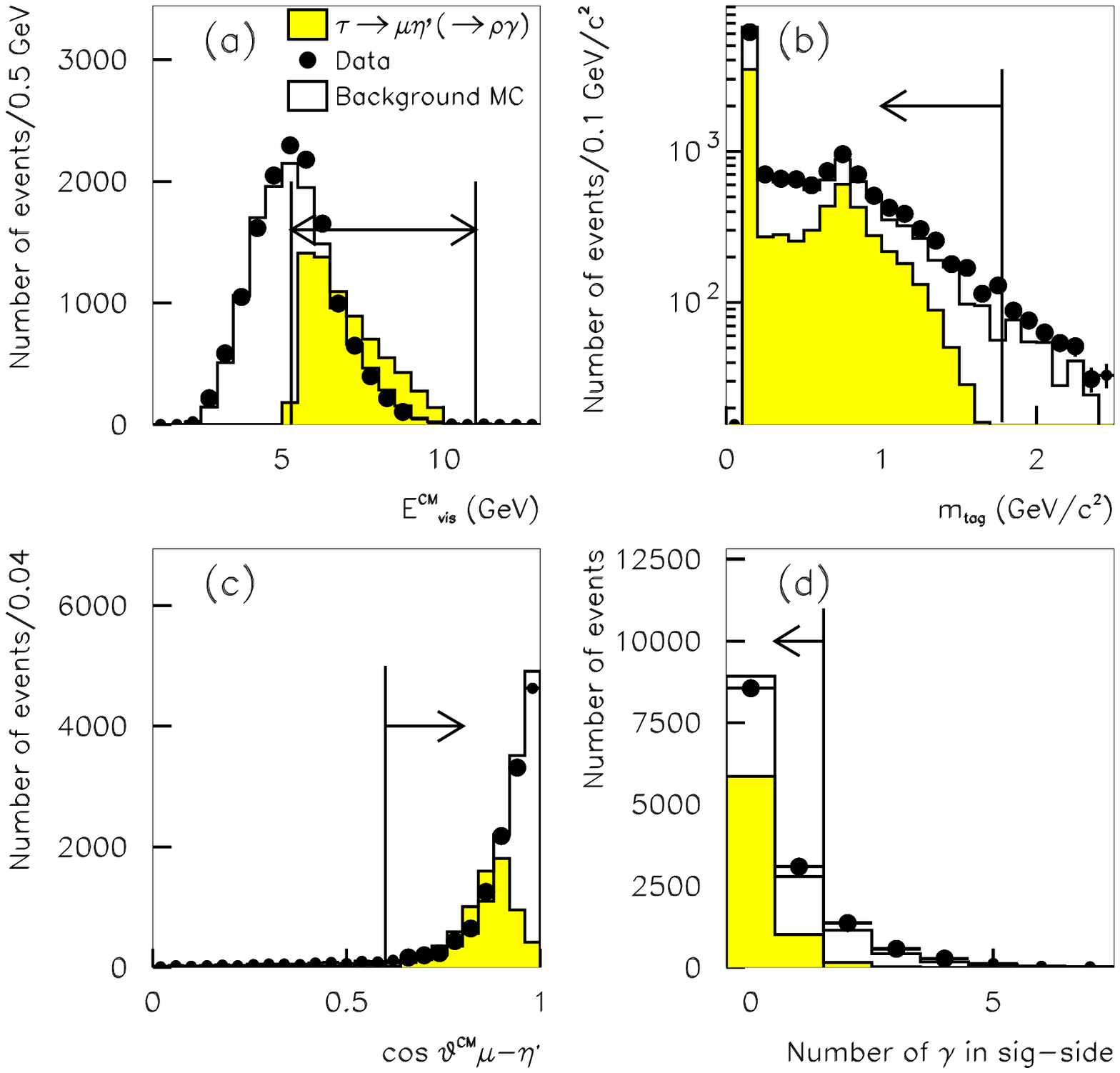}}
 \vspace*{-0.5cm}
\caption{
 {Kinematic distributions used in the event selection
 after $\eta'\to\rho\gamma$ selection, muon {identification and
 missing particles requirements:}
 (a) the total  energy in the CM frame;
 (a) the invariant mass {on} the tag side;
 (c) the opening angle between the missing particle and
 tag-side track in the CM frame;
 (d) the number of {photons} {on} the signal side.
 While the signal MC 
 {($\tau^-\to\mu^-\eta'(\to\rho\gamma)$)} 
 distribution is normalized arbitrarily, 
 {the data and background MC} are normalized to the same luminosity.
 {Selected regions are indicated  
 by arrows from the marked cut {boundaries.}} }
}
\label{fig:muetap_rhogam_dist}
\end{figure}


To suppress 
{fake candidates} 
from 
beam background
and
ISR,
we  {require} the photon energy to
be greater than
0.25 GeV for the barrel 
and  
0.40 GeV for the forward region.
{Similar to the $\eta$ mode,}
we require 
{the following:}
5.29 GeV $< E^{\mbox{\rm{\tiny{CM}}}}_{\rm{vis}} < $11.0 GeV, 
$m_{\rm {tag}} < $ 1.777 GeV/$c^2$,
0.50 $< \cos \theta^{\mbox{\rm \tiny CM}}_{\rm \ell-\eta'}$
(see Fig.~\ref{fig:muetap_rhogam_dist} (a), (b) and (c)).
The requirement on
the number of the photon candidates for the signal
{is }
$n^{\rm{SIG}}\leq 1$,
(see Fig.~\ref{fig:muetap_rhogam_dist} (d)).
We apply the following requirements 
$p_{\rm miss} > -5m^2_{\rm miss}-0.2$
and 
$p_{\rm miss} > 2m^2_{\rm miss}-0.3$ for  {hadronic tags,}
and 
$p_{\rm miss} > -8m^2_{\rm miss}-0.2$
and 
$p_{\rm miss} > 1.2m^2_{\rm miss}-0.3$ for {leptonic tags,}
respectively.
{Following all the selection criteria,}
the signal detection efficiencies 
for {the} electron and
muon modes are
{5.28\% and 6.00\%, 
respectively.
}

\subsection{Event selection for the
$\eta'\to\eta(\to\gamma\gamma)\pi^+\pi^-$ mode}

For $\eta$ meson reconstruction,
we apply
the same {selection criteria} 
as in {the} 
$\tau\to\ell\eta(\to\gamma\gamma)$ analysis.
{The $\eta\to\gamma\gamma$ mass window is chosen to be
0.515 GeV/$c^2$ $< m_{\gamma\gamma} < 0.570$ GeV/$c^2$
and we reject photons from $\pi^0$ decay.}
{Next, we reconstruct  $\eta'$ {candidates} 
using an $\eta$ candidate and two oppositely charged tracks 
consistent with being {pions.}}
We require $P(e) < 0.1$ \text{for both tracks} 
{in} the $\eta'$ candidate.
The $\eta'$ mass window is chosen to be
$0.920$ GeV/$c^2$ $< m_{\eta\pi^+\pi^-} < 0.980$ GeV/$c^2$,
{which corresponds to $\pm$3.0$\sigma$.}

We apply the same cuts on 
$\cos \theta^{\mbox{\rm \tiny CM}}_{\rm thrust-miss}$,
$E^{\mbox{\rm{\tiny{CM}}}}_{\rm{vis}}$,
{$\cos \theta^{\mbox{\rm \tiny CM}}_{\rm \ell-\eta'}$,}
the invariant mass {on} the tag side, 
the number of photons {on} the signal side
and {$m_{\rm{miss}}^2$ vs. $p_{\rm{miss}}$} cut
as {on} {the} 
$\eta'\to\rho\gamma$ analysis.
We also impose the following requirements: 
$p_{\rm miss} > -4m^2_{\rm miss}-0.8$
and 
$p_{\rm miss} > 2.5m^2_{\rm miss}-0.2$ for {hadronic tags,}
and 
$p_{\rm miss} > -3m^2_{\rm miss}$
and 
$p_{\rm miss} > 1.5m^2_{\rm miss}-0.5$ for {leptonic tags,}
respectively, 
where $p_{{\rm tag}}$ is in GeV/$c$.
{Following all the selection criteria,}
the signal detection efficiencies 
for {the} electron and
muon modes are
{4.75\% and  5.47\%, 
respectively.}

\subsection{Event selection for the $\pi^0(\to\gamma\gamma$) modes}

The $\pi^0$ {candidate}
is required to satisfy the condition 
0.115 GeV/$c^2$ $< M_{\gamma\gamma} < $ 0.152 GeV/$c^2$
with $p_{\pi^0} > 0.1$ GeV/c 
{on} the signal side.  
We {require} 
the same cuts as for the $\eta(\to\gamma\gamma)$ mode
since the final {state} is {the} same.
We change 
{the photon energy thresholds} 
for {the} $\pi^0$ candidate 
$E_{\gamma1} >$ 0.90 GeV,
$E_{\gamma2} >$ 0.20 GeV
and $p_{\ell} >$ 1.5 GeV/c
compared to $\tau\to\ell\eta(\to\gamma\gamma)$
(shown in Fig.~\ref{fig:cut_pi0} (a), (b) and (c)). 
Furthermore, 
the opening angle between $\ell$ and
$\pi^0$ {in} the CM system ($\cos \theta^{CM}_{\ell\pi^0} $) 
{should be in the range}
{0.5 $< \cos \theta^{CM}_{\ell\pi^0} <$ 0.8}
(shown in Fig.~\ref{fig:cut_pi0} (d)). 
{Following all the selection criteria,}
the signal detection efficiencies 
for {the} electron and
muon modes are
{4.35\% and 5.03\%, 
respectively.}

\begin{figure}
\begin{center}
       \resizebox{0.8\textwidth}{0.8\textwidth}{\includegraphics
        {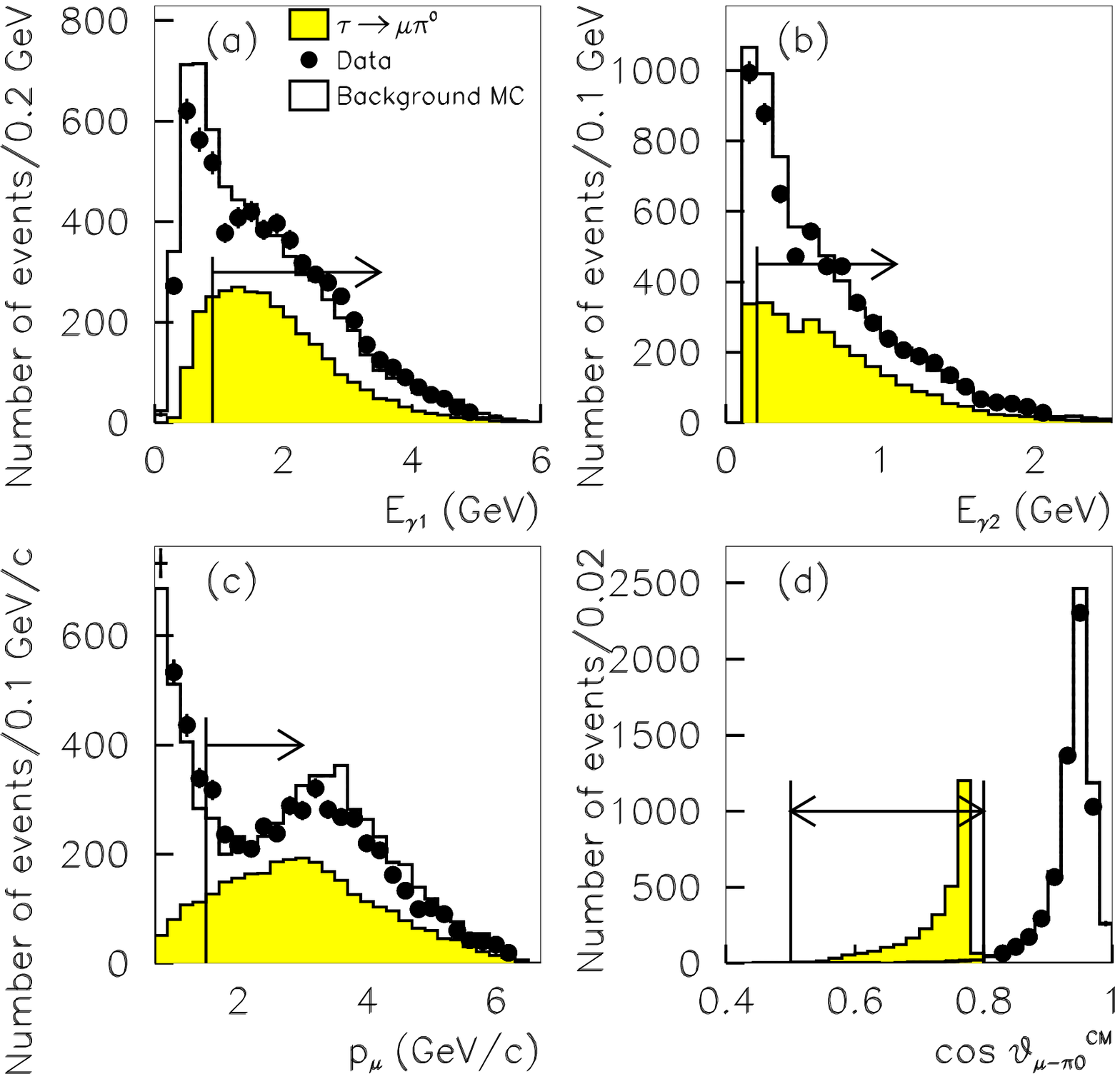}}
 \vspace*{-0.5cm}
 \caption{ 
 {Kinematic distributions used in the event selection:
 (a) higher energy and
 (b) lower energy of a  photon from the $\pi^0$ candidate 
  ($E_{\gamma1}$ and $E_{\gamma2}$);
 (c) momentum of a muon ($p_{\mu}$);
 (d) the cosine of the opening angle between the muon and
 $\pi^0$ in the CM frame ($\cos \theta_{\mu-\pi^0}^{\rm CM}$).
 While the signal MC {($\tau^-\to\mu^-\pi^0$)}
 distribution is normalized arbitrarily, 
 {the data and background MC} are normalized to the same luminosity.
 {Selected regions are indicated  
 by arrows from the marked cut {boundaries.}}} 
}
\label{fig:cut_pi0}
\end{center}
\end{figure}

\section{Signal Region and Background Estimation}

Signal candidates are examined in the two-dimensional
{plots}
of the $\ell^-M^0$ invariant
mass, $M_{\rm {inv}}$, and 
the difference of their energy from the
beam energy in the CM system, $\Delta E$.
A signal event should have $M_{\rm {inv}}$
close to the $\tau$-lepton mass
and
$\Delta E$ close to {zero.}
{For all modes,
the $M_{\rm {inv}}$ and $\Delta E$  resolutions} are parameterized
from the MC distributions 
{{with  asymmetric Gaussian shapes}
to account for initial state radiation and ECL energy leakage for photons.
The resolutions in
$M_{\rm inv}$ and $\Delta E$ are 
{given}
in Table~\ref{tbl:reso_del_e_m}.

\begin{table}
\begin{center}
\begin{tabular}{c|cc|cc} \hline \hline 
Mode
& $\sigma^{\rm{high}}_{M_{\rm{inv}}}$ (MeV/$c^2$)  
& $\sigma^{\rm{low}}_{M_{\rm{inv}}}$ (MeV/$c^2$)
& $\sigma^{\rm{high}}_{\Delta E}$ (MeV)      
&  $\sigma^{\rm{low}}_{\Delta E}$ (MeV)
 \\ \hline
$\mu\eta(\to\gamma\gamma)$ 
& 14.7 &  19.4 & 30.3  & 61.4  \\  
$\mu\eta(\to\pi^+\pi^-\pi^0)$ 
& 7.2  & 8.5 & 18.5 & 36.4 \\  \hline
$e\eta(\to\gamma\gamma)$ 
& 14.0  &  19.8 & 37.3 & 62.4  \\  
$e\eta(\to\pi^+\pi^-\pi^0)$ 
 & 7.6 &  9.3 & 19.4 & 41.8 \\  \hline\hline
$\mu\eta'(\to\rho\gamma)$
 & 7.8  &  9.0 & 16.8  & 34.1 \\

$\mu\eta'(\to\eta\pi^+\pi^-)$
& 11.2  & 19.1 & 27.1 & 53.5  \\  \hline

$e\eta'(\to\rho\gamma)$
&  9.2  &  10.4 & 19.6 & 40.0  \\

$e\eta'(\to\eta\pi^+\pi^-)$
 & 10.3  &  21.9 & 26.1 & 59.4  \\  \hline\hline

$\mu\pi^0$
& 14.9  &  19.1 & 33.8 & {63.0} \\
$e\pi^0$
 & 12.7  & 23.1 & 35.6 & 64.6  \\  \hline\hline

\end{tabular}
\caption{Summary {of} $M_{\rm inv}$ (MeV/$c^2$) and 
$\Delta E$ 
{resolutions} (MeV)}
\label{tbl:reso_del_e_m}
\end{center}
\end{table}

{To evaluate the branching fraction, 
we use  elliptical regions, which {contain} 90\% 
of the MC signal that {satisfies} all cuts.
We find that an elliptical signal region gives better 
sensitivity
than {a} rectangular one.
The signal regions are shown in Fig.~\ref{fig:openbox};
the corresponding signal {efficiencies} are given in Table~\ref{tbl:eff}.}

We blind {the signal region}
{so as not to bias our choice of selection criteria.}
Figures~\ref{fig:openbox} 
and ~\ref{fig:openbox2} show 
scatter-plots
for data and signal MC samples 
distributed over $\pm 10\sigma$
in the $M_{\rm{inv}}-\Delta E$ plane.
{As there are few remaining MC background events 
in the signal ellipse, 
we estimate the background contribution 
using the $M_{\rm{inv}}$  sideband regions.
Extrapolation to the signal region assumes that
the background distribution  is
flat along the $M_{\rm inv}$ axis.}
We then estimate {the} expected number of the background events in the signal
region for each  mode
using  the number of 
{data events observed} 
in the sideband region  
inside 
{the} horizontal lines {but}
excluding the signal region
as shown in Fig.~\ref{fig:openbox} and \ref{fig:openbox2}.
The numbers of background events in the 90\% elliptical signal region 
{are} {also}
shown in Table~\ref{tbl:eff}.

\begin{table}
\begin{center}
\begin{tabular}{|c|c|c|c|c|c|c|}\hline \hline
Mode & ${\cal{B}}_{M^0}$ & $\varepsilon$ & 
 $b_0$  &  $s$ & Total Sys. & $s_{90}$ \\ \hline
$\tau\to\mu\eta(\to\gamma\gamma)$ & 0.3943 & 6.42\% & 0.40$\pm$0.29 & 0 &
 7.1\% & 2.1 \\
$\tau\to\mu\eta(\to\pi^+\pi^-\pi^0)$ & 0.226 & 6.84\% & 
 0.24$\pm$0.24 & 0 & 5.6\% & 2.2  \\ \hline
$\tau\to e\eta(\to\gamma\gamma)$ & 0.3943 & 4.57\% & 0.25$\pm$0.25 
 & 0 & 7.1\% & 2.2 \\
$\tau\to e\eta(\to\pi^+\pi^-\pi^0)$ & 0.226 & 4.72\% & 0.53$\pm$0.53
 & 0 & 5.6 \% & 2.0   \\ \hline\hline
$\tau\to\mu\eta'(\to\rho\gamma)$ & 0.295$\times$1.0 & 5.40\% &
 0.23$\pm$0.23 & 0 & 6.8\% &  2.2 \\
$\tau\to\mu\eta'(\to\eta\pi^+\pi^-)$ & 0.443$\times$0.3943 & 4.92\%&
 0.0${}^{+0.23}_{-0.0}$ & 0 & 8.9\% & 2.5  \\ \hline
$\tau\to e\eta'(\to\rho\gamma)$ & 0.295$\times$1.0 & 4.76\% & 0.0${}^{+0.33}_{-0.0}$
 & 0 & 6.8\% & 2.5 \\
$\tau\to e\eta'(\to\eta\pi^+\pi^-)$ & 0.443$\times$0.3943 & 4.27\% & 0.0${}^{+0.24}_{-0.0}$
 & 0 & 8.9\% & 2.5  \\ \hline\hline
$\tau\to\mu\pi^0(\to\gamma\gamma)$ & 0.98798 & 4.53\% & 0.58$\pm$0.34 & 1
 & 4.5\% & 3.8\\
$\tau\to e\pi^0(\to\gamma\gamma)$ & 0.98798 & 3.93\% & 0.20$\pm$0.20
 & 0 & 4.5\% & 2.2 \\
\hline\hline
\end{tabular}
\caption{Results of the final event selection for 
the individual modes:
${\cal{B}}_{M^0}$ is the branching fraction for the $M^0$ decay;
$b_0$ and $s$ are the number of expected background and 
observed events in the signal region, respectively;
``Total sys.'' means the total systematic uncertainty; 
$s_{90\%}$ is the upper limit on the number of signal events
including systematic uncertainties.
}
\label{tbl:eff}
\end{center}
\end{table}

%
%
%


Systematic uncertainties for  $M^0$ reconstruction are 
3.0\%, 4.0\%, 4.0\%, 5.0\% and 3.0\%
for 
$\eta\to\gamma\gamma$, $\eta\to\pi^+\pi^-\pi^0$,
$\eta'\to\rho\gamma$, 
{$\eta'\to\eta\pi^+\pi^-$} and 
$\pi^0\to\gamma\gamma$,
respectively. 
Furthermore,
the uncertainties due to the branching ratios of the $M^0$ meson are
{0.7\%,} 1.8\%, 3.4\% and 3.5\% 
for $\eta\to\gamma\gamma$,  $\eta\to\pi^+\pi^-\pi^0$,
$\eta'\to\rho\gamma$ and  $\eta'\to\eta\pi^+\pi^-$, respectively
~\cite{PDG}. 
For the $\pi^0$ veto  we take a 5.5\% uncertainty  
{for} $\eta\to\gamma\gamma$
while a 2.8\% uncertainty  
is assigned {to} 
the $\eta'\to\rho\gamma$ mode. 
The uncertainties {in the} trigger (0.5$-$1.0\%), tracking (2.0\%), 
{lepton identification (2.0\%),}
MC statistics (1.0$-$1.5\%), luminosity (1.4\%) are also considered. 
{All these uncertainties are added in quadrature, 
and the total systematic uncertainties are
shown in Table~\ref{tbl:eff}.}  

While the angular distribution of signal $\tau$ decays is 
initially assumed to be uniform {in this analysis},
it is sensitive to the lepton flavor violating interaction
structure~\cite{LFV}.
The spin correlation 
between the $\tau$ lepton {on} the signal and that {on}  the tag side
must be considered.
A possible nonuniformity was taken into account by comparing
the uniform case with {MC's}
assuming $V-A$ and $V+A$ interactions,
which result in the maximum possible variations.
No statistically significant difference in 
the $M_{\rm inv}$ -- $\Delta{E}$
distribution or the efficiencies is found {compared to}
the case of the uniform distribution.
Therefore,
systematic uncertainties due to these effects 
are neglected in {the} upper limit evaluation.

We open the blind and find no data events  
in the {blinded} region after event selection.
{Only in the case of 
$\tau\to\mu\pi^0(\to\gamma\gamma)$,
one event is found in the elliptical region.
Since no statistically significant excess of data over
the expected background in the signal region is observed,} 
we set upper limits for branching fractions.
The upper limit {on the number of} signal events 
at the 90\% confidence level (C.L.) $s_{90}$ 
including  systematic uncertainty 
is obtained 
with the use of the Feldman-Cousins method~\cite{cite:FC}
calculated 
by the POLE program without conditioning \cite{pole}.   
The upper limit {on} the branching fraction ($\cal{B}$) is then calculated as 
\begin{equation}
{\cal{B}}(\tau^-\to\ell^- M^0) <
\displaystyle{\frac{s_{90}}{2N_{\tau\tau}\varepsilon{{\cal{B}}_{M^0}}}}
\end{equation}
where
{${\cal B}_{M^0}$ {is taken} from 
{PDG}~\cite{PDG}}
and
{$N_{\tau\tau} =  357.7\times 10^6$
is
{the number of $\tau-$pairs
{produced} in
401 fb${}^{-1}$ of data.
We obtain $N_{\tau\tau}$
using
$\sigma_{\tau\tau} = 0.892 \pm 0.002$ nb,
the $e^+e^- \rightarrow \tau^+\tau^-$ cross section
at the $\Upsilon(4S)$ resonance
calculated by KKMC~\cite{KKMC}.}}
The upper limits for the branching fractions are
{in the range}
${\cal{B}}(\tau^-\rightarrow \ell^- M^0) < (6.5-16)\times 10^{-8}$ 
at the 90\% confidence level, respectively. 
{A} summary {of} the upper limits is 
{given} in Table~\ref{tbl:results}.
These results improve the previously published limits~\cite{cite:leta} 
by factors of 2.3$-$6.4.

\begin{table}
\begin{center}
\begin{tabular}{|l|l||c|}\hline\hline
Mode & $M^0$ subdecay mode & Upper limit of $\cal B$ at 90\%C.L. \\\hline
$\tau^-\to \mu^-\eta$ & $\eta\to\gamma\gamma$ & 1.2$\times 10^{-7}$ \\
                  & $\eta\to\pi^+\pi^-\pi^0$ & 2.0$\times 10^{-7}$ \\
                  & Combined  & 6.5$\times 10^{-8}$ \\\hline \hline 
$\tau^-\to e^-\eta$ & $\eta\to\gamma\gamma$ & 1.7$\times 10^{-7}$ \\
                  & $\eta\to\pi^+\pi^-\pi^0$ & 2.6$\times 10^{-7}$ \\
                  & Combined  & 9.2$\times 10^{-8}$ \\\hline \hline 
$\tau^-\to \mu^-\eta'$ & $\eta\to\rho\gamma$ & 1.9$\times 10^{-7}$ \\
                  & $\eta\to\eta\pi^+\pi^-$ & 4.1$\times 10^{-7}$ \\
                  & Combined  & 1.3$\times 10^{-7}$ \\\hline \hline 
$\tau^-\to e^-\eta'$ & $\eta\to\rho\gamma$ & 2.5$\times 10^{-7}$ \\
                  & $\eta\to\eta\pi^+\pi^-$ & 4.7$\times 10^{-7}$ \\
                  & Combined  & 1.6$\times 10^{-7}$ \\\hline \hline 
$\tau^-\to \mu^-\pi^0$ & $\pi^0\to\gamma\gamma$ & 1.2$\times 10^{-7}$ \\\hline \hline 
$\tau^-\to e^-\pi^0$ & $\pi^0\to\gamma\gamma$ & 8.0$\times 10^{-8}$ \\\hline \hline 
\end{tabular}
\caption{Summary {of} upper limits {on} $\cal B$ at 90\%C.L.}
\label{tbl:results}
\end{center}
\end{table}

\begin{figure}
\begin{center}
 \resizebox{0.3\textwidth}{0.3\textwidth}{\includegraphics
 {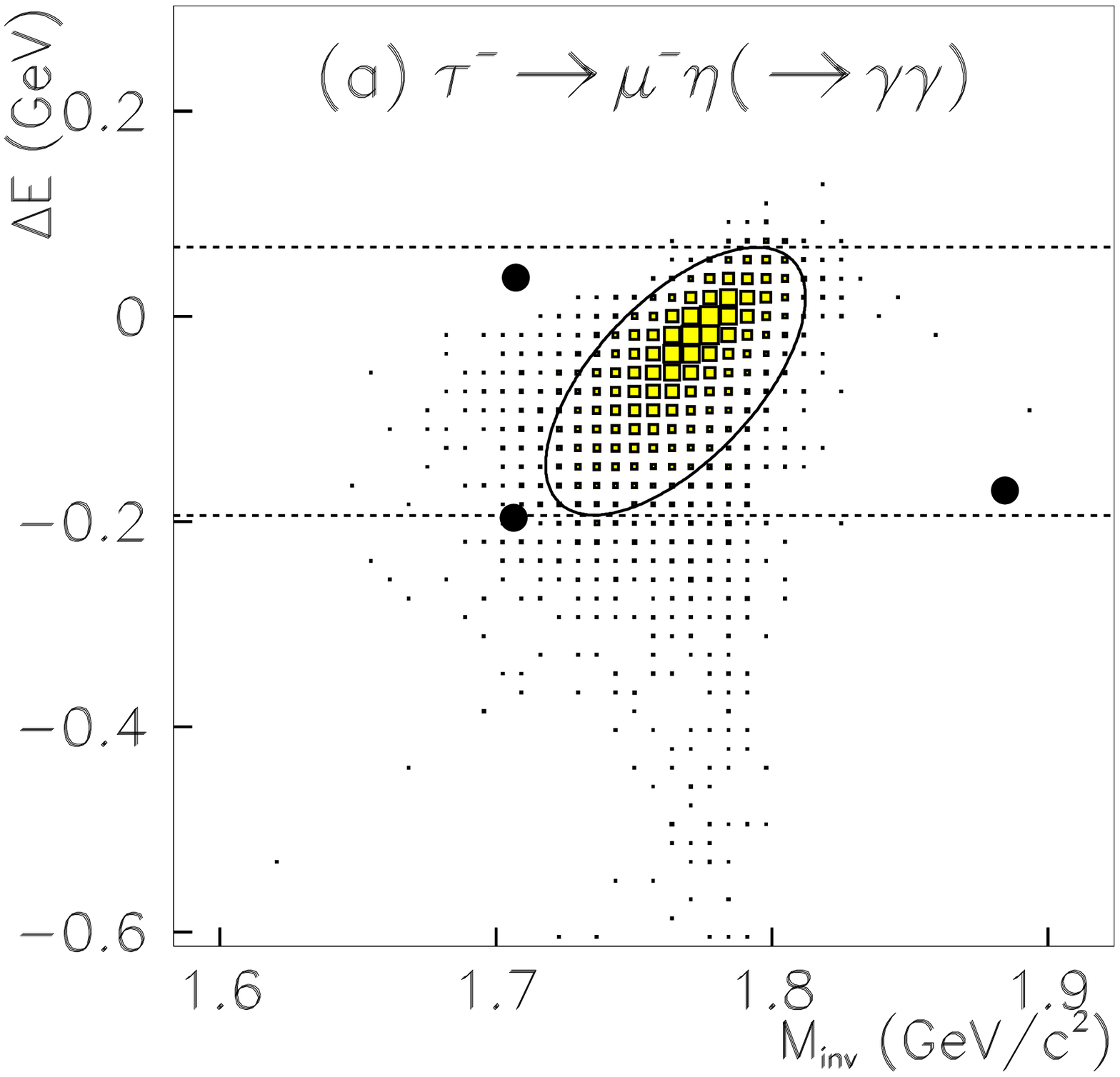}}
 \resizebox{0.3\textwidth}{0.3\textwidth}{\includegraphics
 {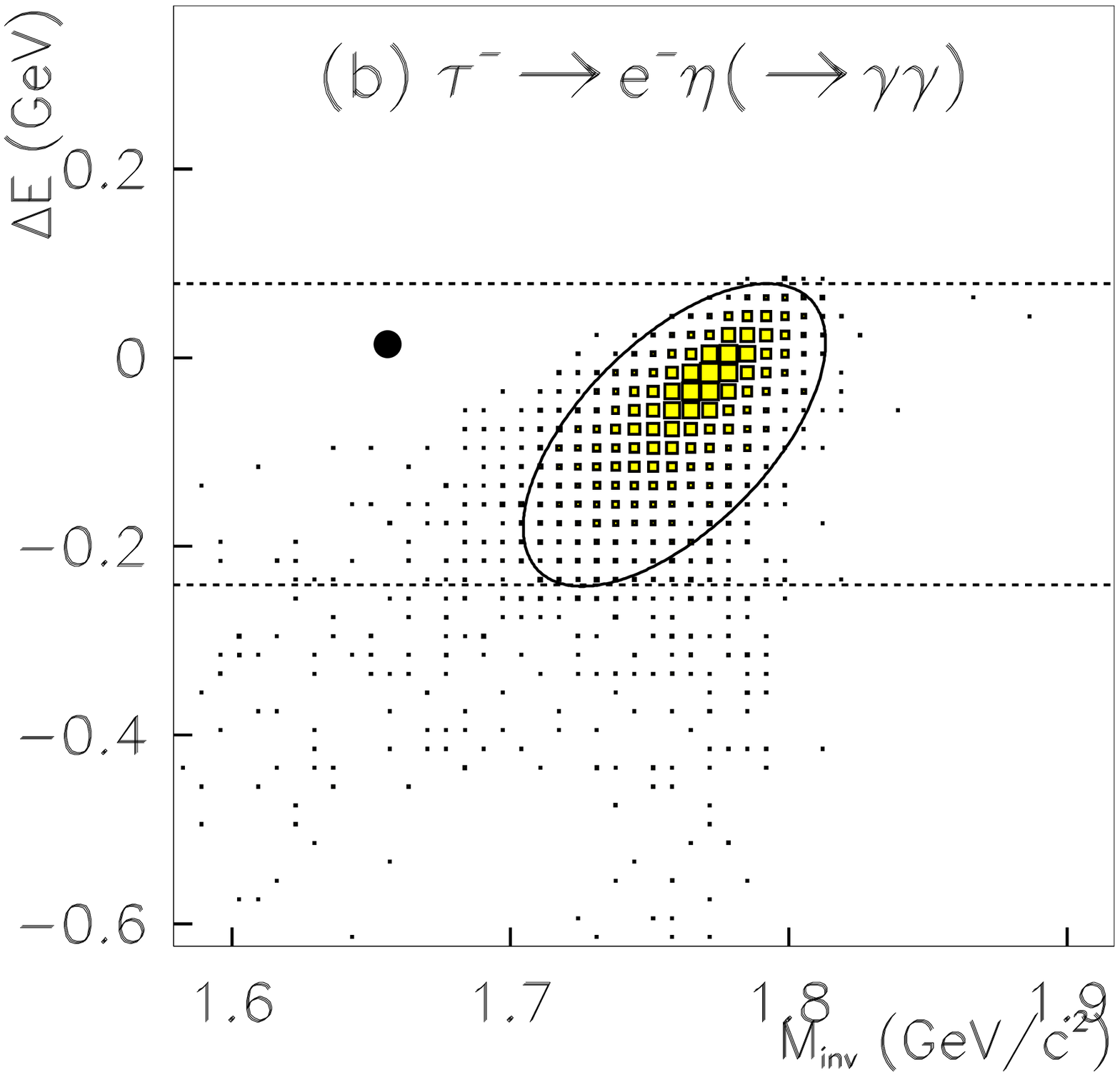}}\\
 \vspace*{-0.5cm} 
 \resizebox{0.3\textwidth}{0.3\textwidth}{\includegraphics
 {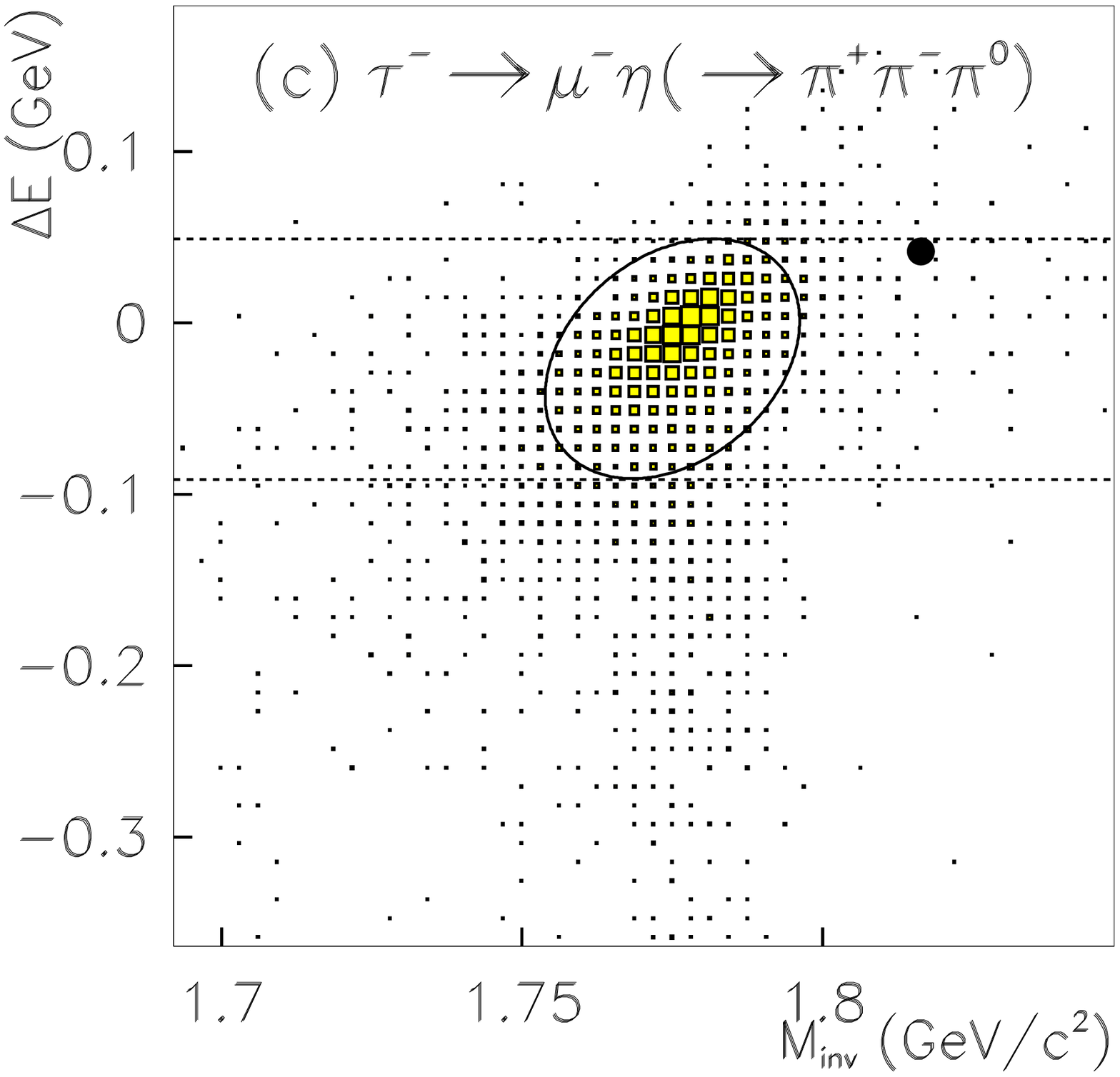}}
 \resizebox{0.3\textwidth}{0.3\textwidth}{\includegraphics
 {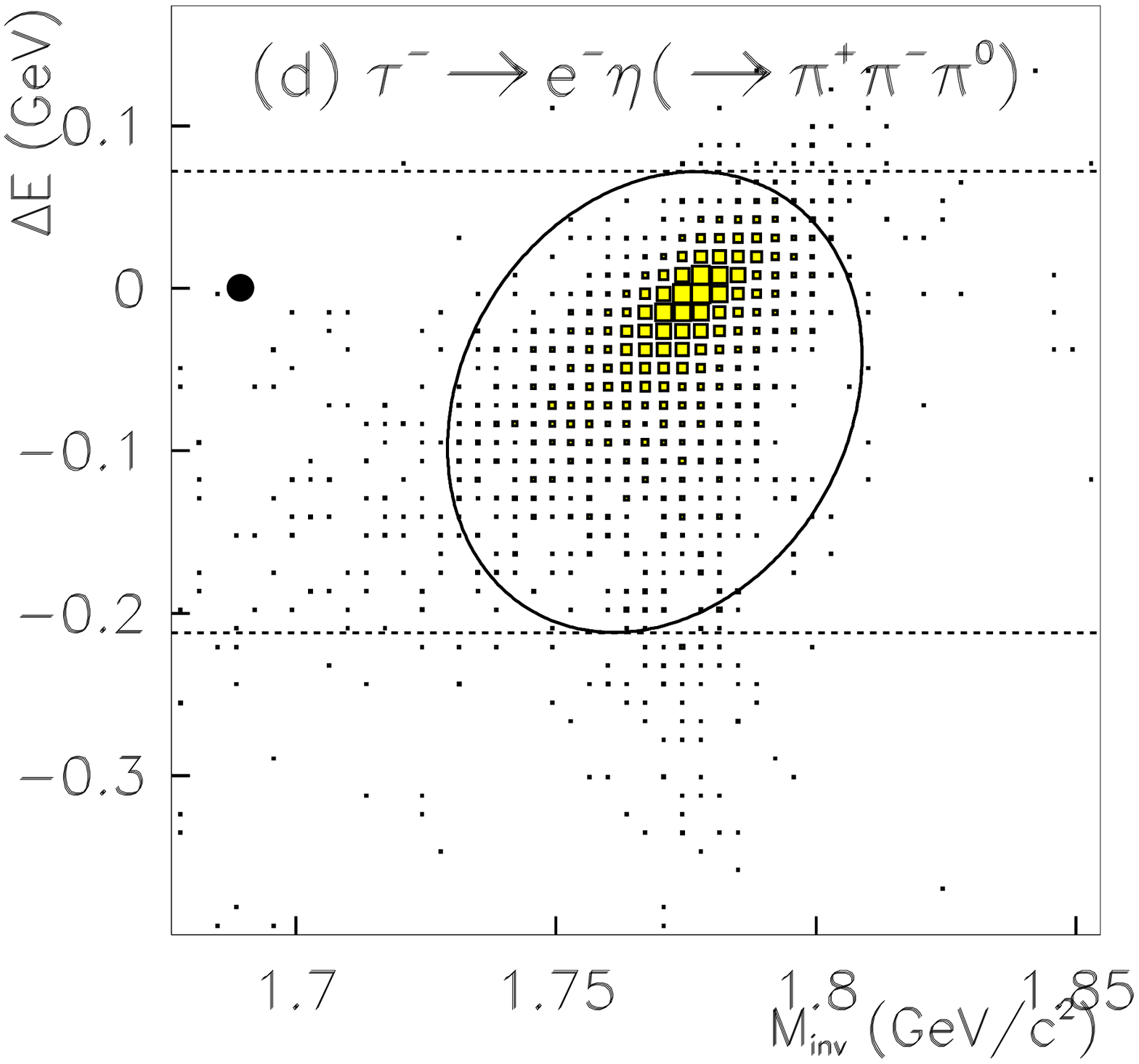}}\\
\vspace*{-0.5cm} 
\caption{
Scatter-plots of data in the
$M_{\rm inv}$ -- $\Delta{E}$ plane:
(a), (b), (c) and (d) correspond to
the $\pm 10 \sigma$ area for
the $\tau^-\rightarrow\mu^-\eta(\to\gamma\gamma)$,
$\tau^-\rightarrow e^-\eta(\to\gamma\gamma)$,
$\tau^-\rightarrow\mu^-\eta(\to\pi^+\pi^-\pi^0)$ and
$\tau^-\rightarrow e^-\eta(\to\pi^+\pi^-\pi^0)$,
modes, respectively.
The filled boxes show the MC signal distribution
with arbitrary normalization.
The elliptical signal region shown by the solid curve
is used for evaluating the signal yield.
The region between the horizontal lines excluding the signal region is
used to estimate the expected background in the elliptical region.
}
\label{fig:openbox}
\end{center}
 \end{figure}

\begin{figure}
\begin{center}
 \resizebox{0.3\textwidth}{0.3\textwidth}{\includegraphics
{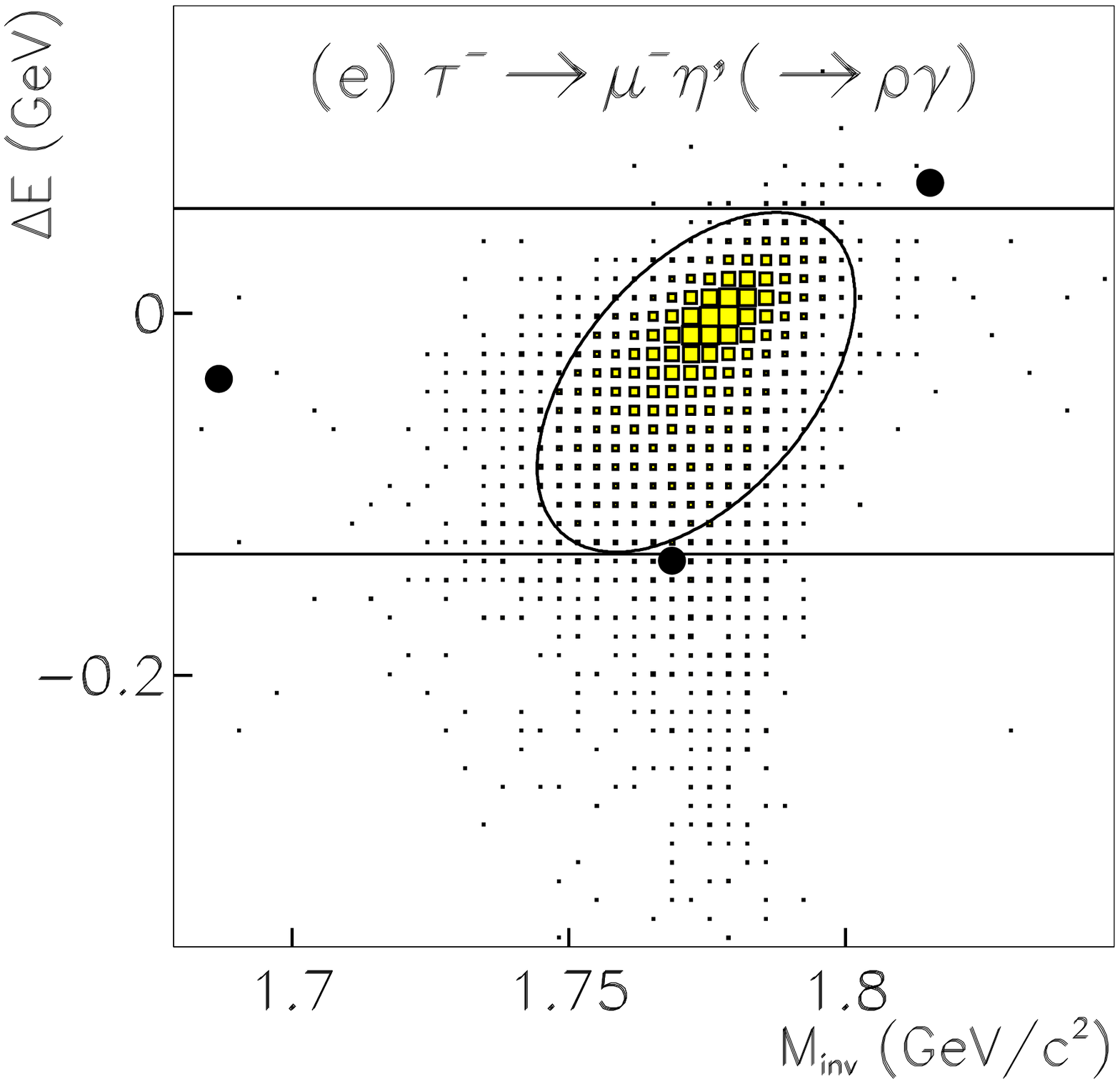}}
 \resizebox{0.3\textwidth}{0.3\textwidth}{\includegraphics
{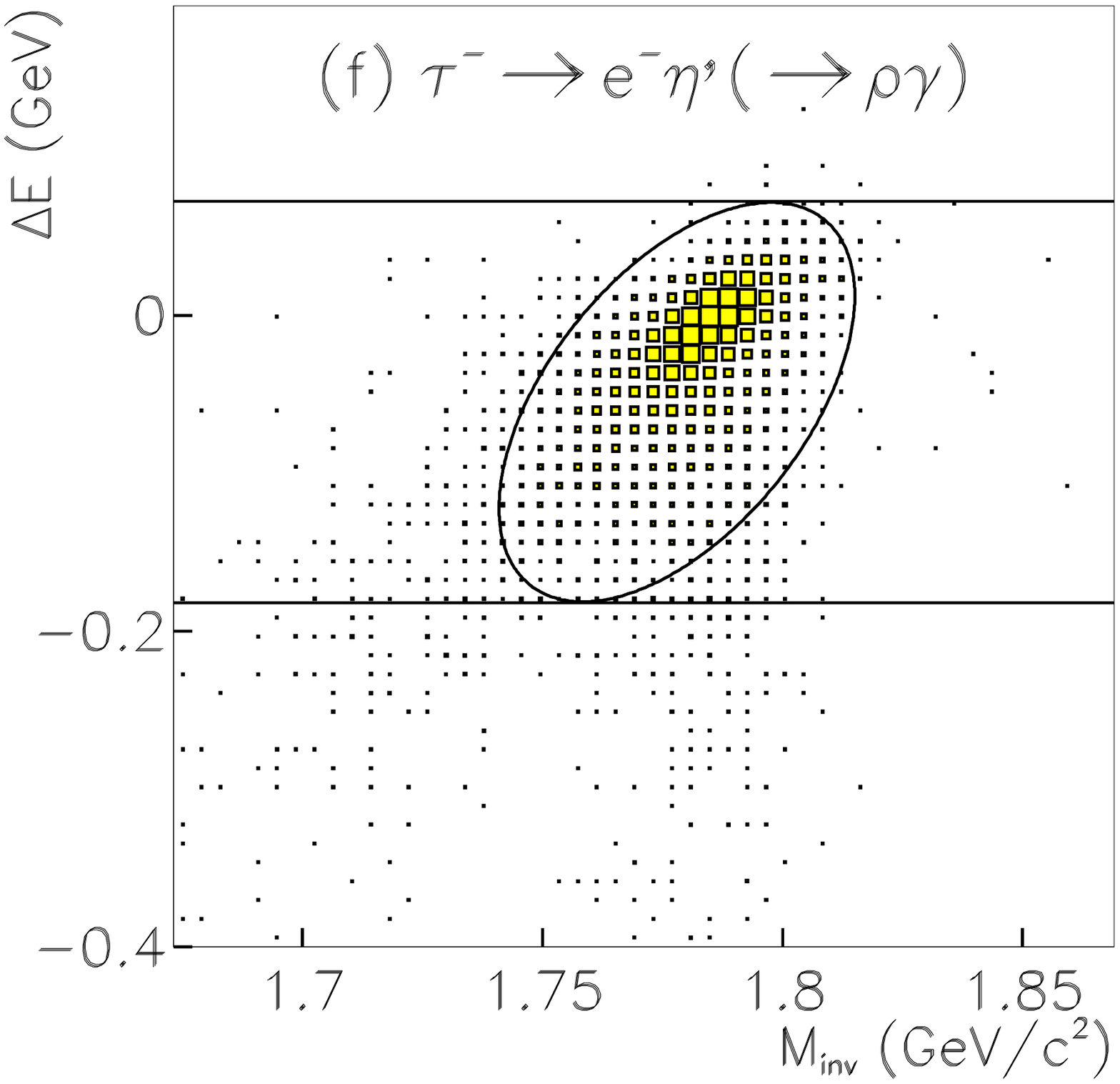}}\\
\vspace*{-0.5cm} 
 \resizebox{0.3\textwidth}{0.3\textwidth}{\includegraphics
{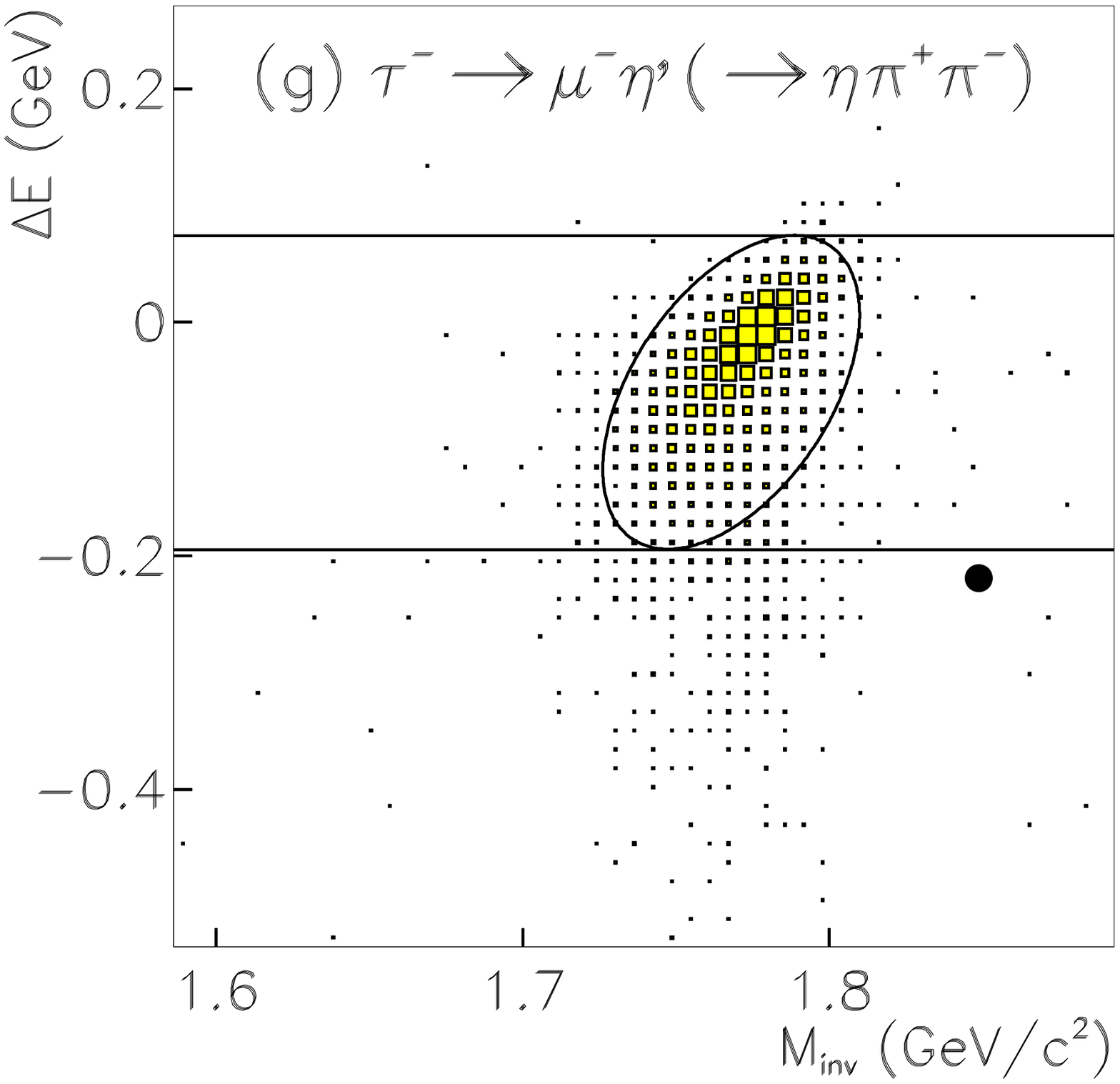}}
 \resizebox{0.3\textwidth}{0.3\textwidth}{\includegraphics
{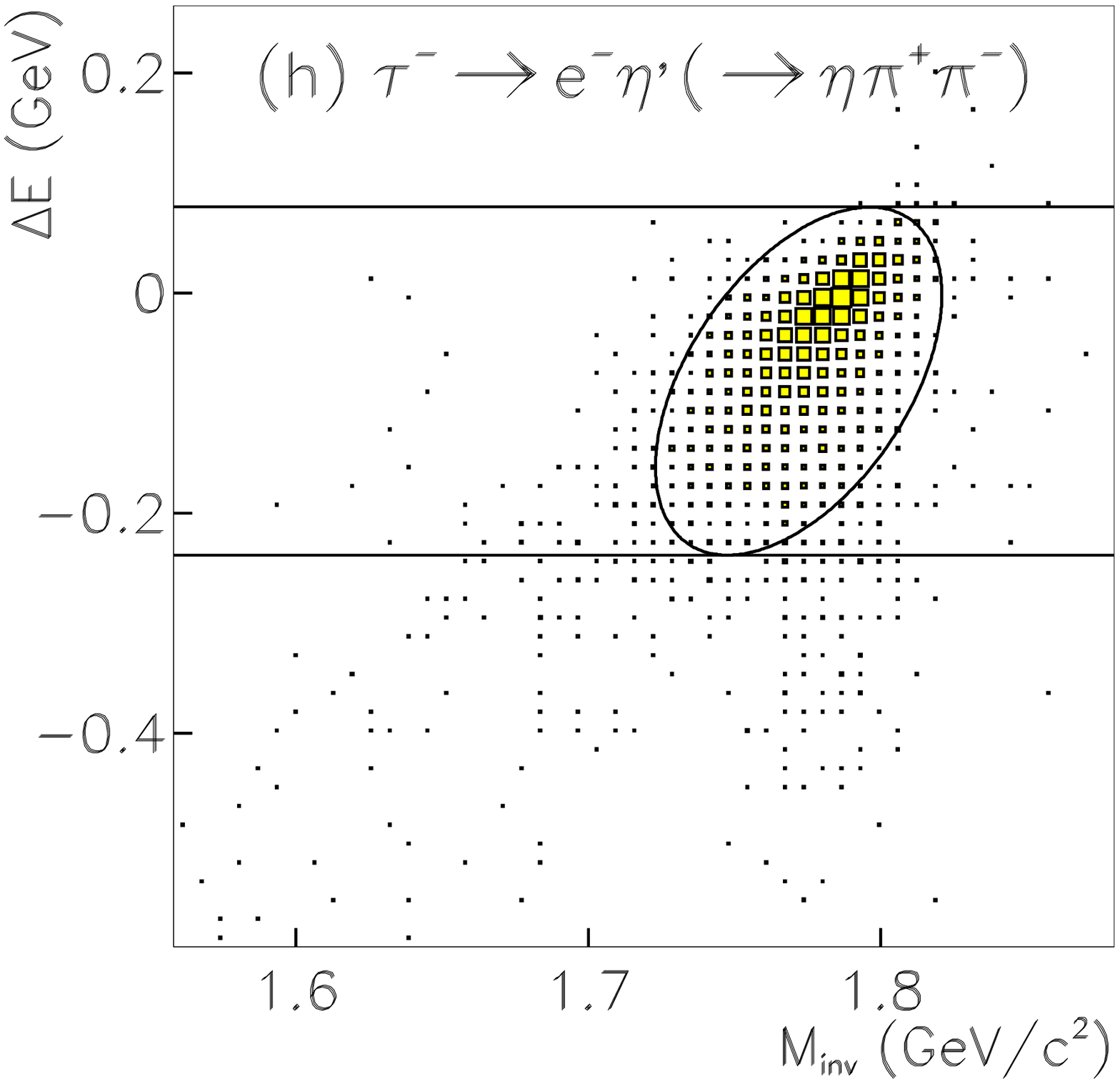}}\\
\vspace*{-0.5cm} 
 \resizebox{0.3\textwidth}{0.3\textwidth}{\includegraphics
{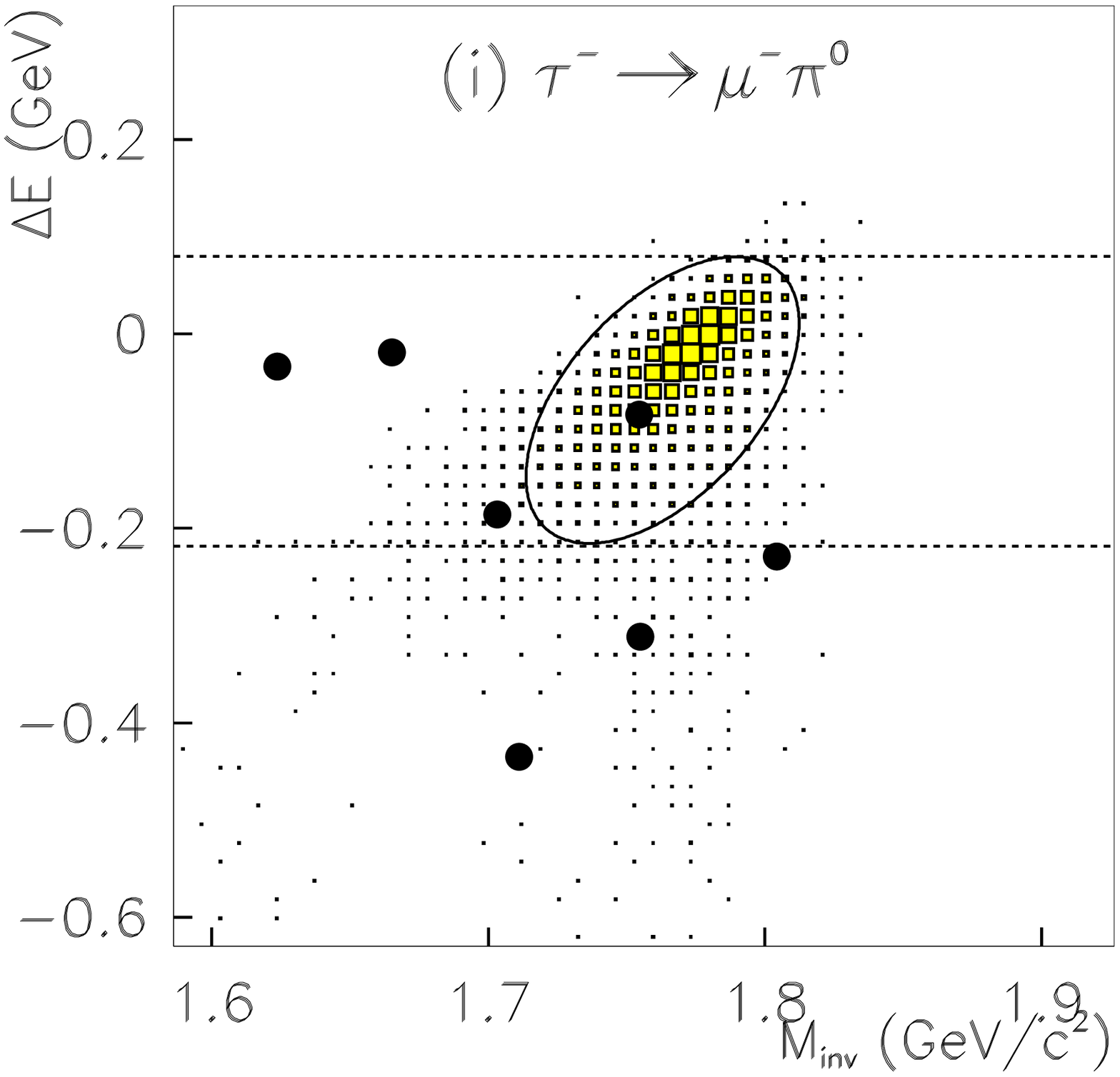}}
 \resizebox{0.3\textwidth}{0.3\textwidth}{\includegraphics
{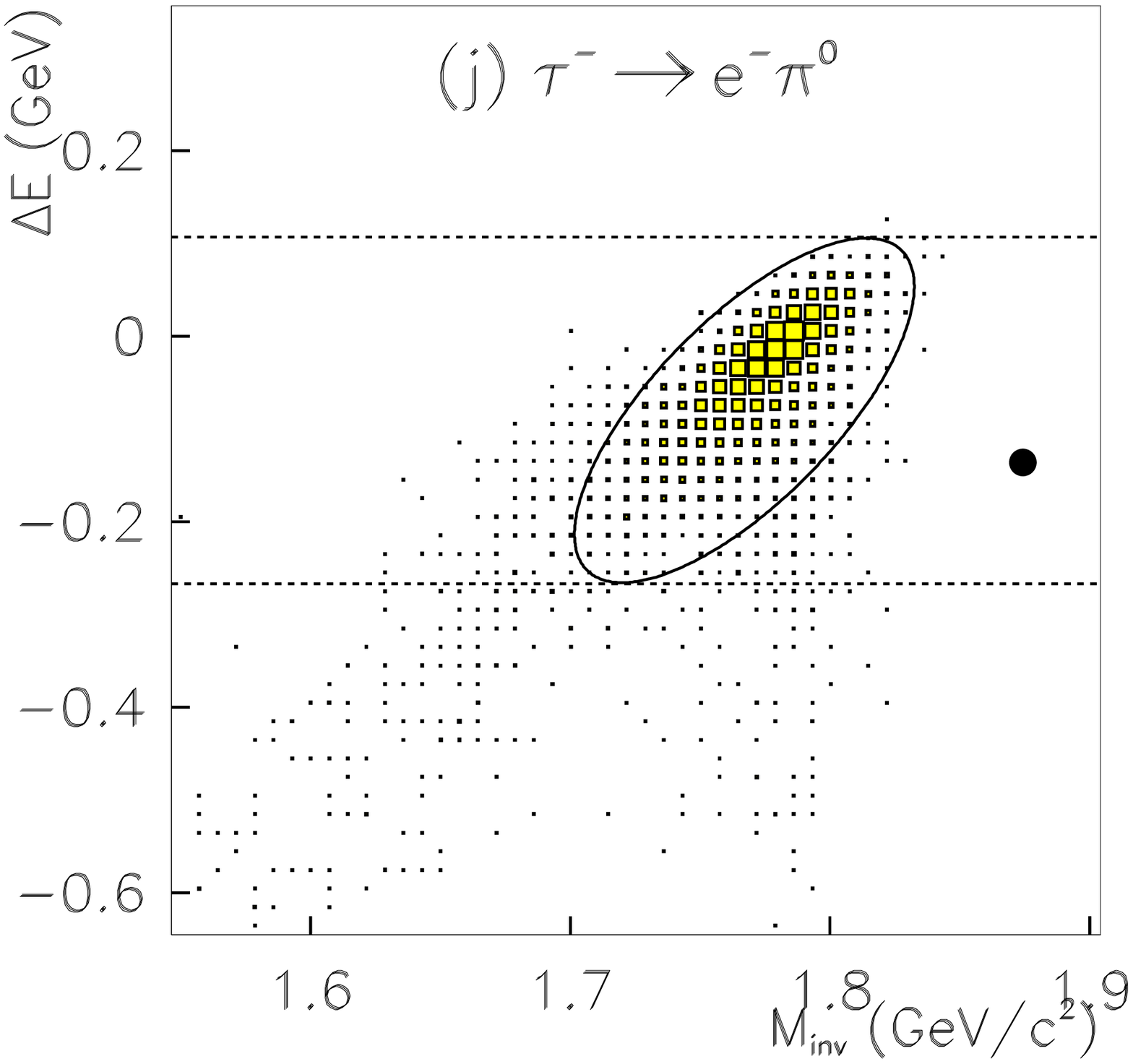}}\\
\caption{
Scatter-plots of data in the
$M_{\rm inv}$ -- $\Delta{E}$ plane:
(e), (f), (g),  (h), (i) and (j) correspond to
the $\pm 10 \sigma$ area for
the $\tau^-\rightarrow\mu^-\eta'(\to\rho\gamma)$,
$\tau^-\rightarrow e^-\eta'(\to\rho\gamma)$,
$\tau^-\rightarrow\mu^-\eta'(\to\eta\pi^+\pi^-)$,
$\tau^-\rightarrow e^-\eta'(\to\eta\pi^+\pi^-)$,
$\tau^-\rightarrow\mu^-\pi^0(\to\gamma\gamma)$ and 
$\tau^-\rightarrow e^-\pi^0(\to\gamma\gamma)$
modes, respectively.
{The data are indicated by the solid circles.}
The filled boxes show the MC signal distribution
with arbitrary normalization.
The elliptical signal region shown by the solid curve
is used for evaluating the signal yield.
The region between the horizontal lines excluding the signal region is
used to estimate the expected background in the elliptical region.
}
\label{fig:openbox2}
\end{center}
\end{figure}

\section{Discussion}

The branching ratio {for} the  $\tau^-\to\mu^-\eta$ mode
is enhanced by Higgs-mediated LFV 
if large mixing between a left-hand
scalar muon and scalar tau in the corresponding SUSY
model occurs~\cite{cite:higgs} and can be written as
\begin{equation}
{\cal{B}}(\tau^-\to\mu^-\eta) =
8.4\times10^{-7}
\displaystyle\left(
\frac{\tan \beta}{60}
\right)^6
\left(
\displaystyle\frac{100 \mbox{GeV}/c^2}{m_A}
\right)^4,
\end{equation}
where $m_A$ is the pseudoscalar Higgs mass
and $\tan \beta$ is the ratio of the vacuum expectation values.
From our upper limit {on}
the branching fraction {for} the $\tau^-\to\mu^-\eta$ decay,
some region of $m_A$ and $\tan \beta$ parameters
can be excluded.
Figure~\ref{fig:mAtanB}
shows the excluded region in the $m_A$ vs $\tan \beta$ plane.
Figure~\ref{fig:mAtanB} also shows
the constraints at 95\% C.L. from the
CDF~\cite{CDF}, $\mbox{D\O}$~\cite{D0} and LEP2 experiments~\cite{LEP}.
The excluded regions 
from the 
CDF, $\mbox{D\O}$ and LEP2 experiments
are
shown
with $\mu > 0$ and
$m^{\rm max}_{h}$.
Our result has a sensitivity competitive
with that 
of the CDF and  $\mbox{D\O}$ experiments,
which searched for 
$p\bar{p}\to\phi b(\bar{b})$ and $\tau^+\tau^-$ events,
where $\phi$ is a neutral Higgs boson in 
the minimal {supersymmetric} standard model~($\phi = h, H$ and $A$).

\begin{figure}
\begin{center}
 \resizebox{0.45\textwidth}{0.45\textwidth}{\includegraphics
{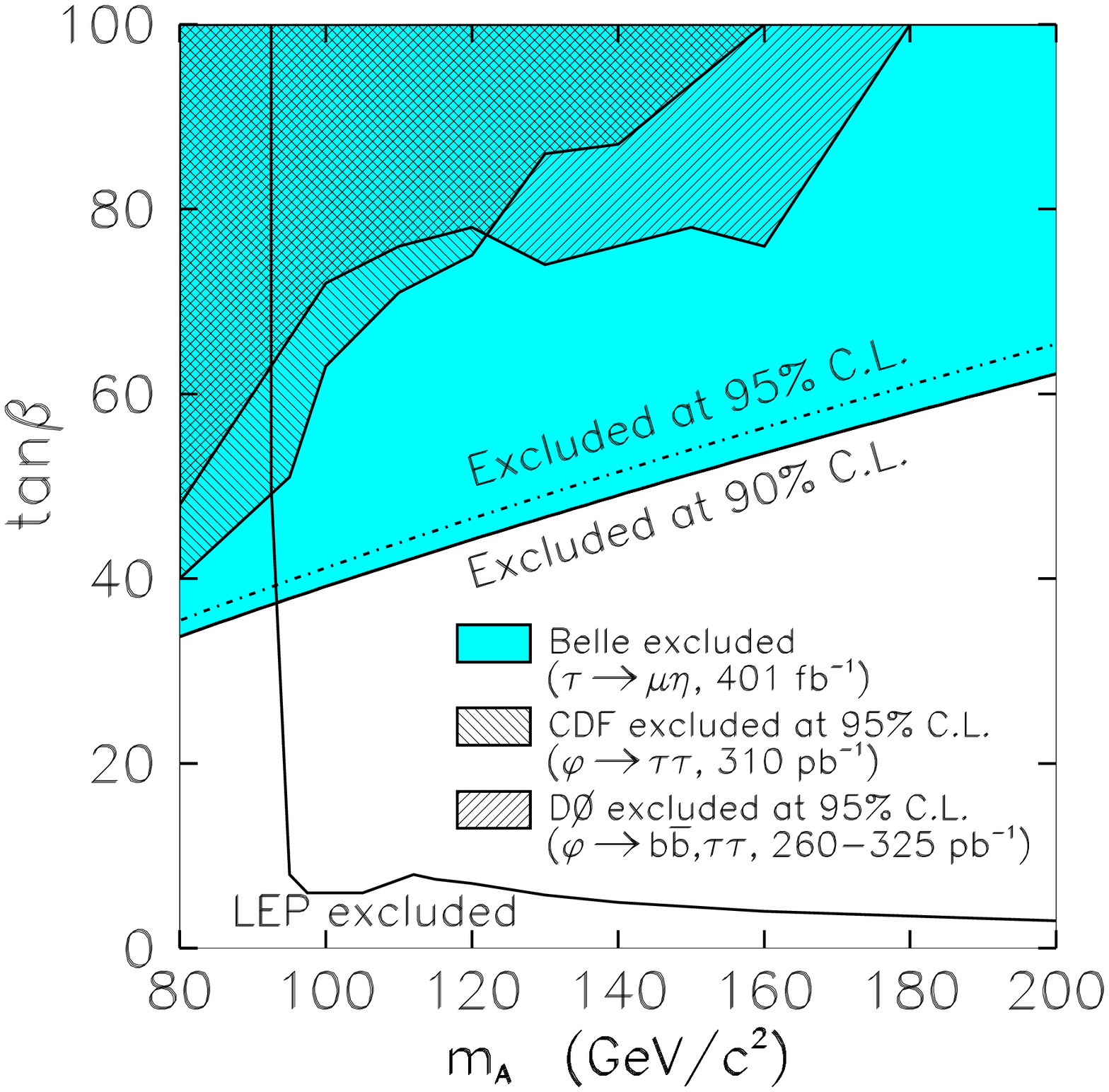}}
 \vspace*{-0.5cm}
\caption{
{The excluded region} 
{in} the $m_A$ vs $\tan \beta$ plane
from our results at 90\% C.L.
and
other experiments at 95\% C.L. from
CDF~\cite{CDF}, $\mbox{D\O}$~\cite{D0}, LEP~\cite{LEP}. 
The excluded regions 
from the
CDF, $\mbox{D\O}$ and LEP2 experiments
are
shown
with $\mu > 0$ and
$m^{\rm max}_{h}$.
}
\label{fig:mAtanB}
\end{center}
\end{figure}

\section{Summary}

We have searched for lepton flavor violating $\tau$ 
{decays} with a pseudoscalar meson 
{($\eta$, $\eta'$ and $\pi^0$)}
using 401 fb$^{-1}$ of data.
No signal is found and
we set the following upper limits of the branching fractions: 
${\cal{B}}(\tau^-\rightarrow e^-\eta) < 9.2\times 10^{-8}$, 
${\cal{B}}(\tau^-\rightarrow \mu^-\eta) < 6.5\times 10^{-8}$,
${\cal{B}}(\tau^-\rightarrow e^-\eta') < 1.6\times 10^{-7}$, 
${\cal{B}}(\tau^-\rightarrow \mu^-\eta') < 1.3\times 10^{-7}$  
${\cal{B}}(\tau^-\rightarrow e^-\pi^0) < 8.0\times 10^{-8}$
and 
${\cal{B}}(\tau^-\rightarrow \mu^-\pi^0) < 1.2\times 10^{-7}$    
at the 90\% confidence level, respectively. 
These results improve the previously published limits 
by factors from 2.3 to 6.4 
{and help to constrain physics beyond the Standard Model.}

\section*{Acknowledgments}

We thank the KEKB group for the excellent operation of the
accelerator, the KEK cryogenics group for the efficient
operation of the solenoid, and the KEK computer group and
the National Institute of Informatics for valuable computing
and Super-SINET network support. We acknowledge support from
the Ministry of Education, Culture, Sports, Science, and
Technology of Japan and the Japan Society for the Promotion
of Science; the Australian Research Council and the
Australian Department of Education, Science and Training;
the National Science Foundation of China and the Knowledge
Innovation Program of the Chinese Academy of Sciencies under
contract No.~10575109 and IHEP-U-503; the Department of
Science and Technology of India;
the BK21 program of the Ministry of Education of Korea,
the CHEP SRC program and Basic Research program
(grant No.~R01-2005-000-10089-0) of the Korea Science and
Engineering Foundation, and the Pure Basic Research Group
program of the Korea Research Foundation;
the Polish State Committee for Scientific Research;
the Ministry of Science and Technology of the Russian
Federation; the Slovenian Research Agency;  the Swiss
National Science Foundation; the National Science Council
and the Ministry of Education of Taiwan; and the U.S.\
Department of Energy.

\newpage

\end{document}